\documentclass[twoside,fleqn]{article}
\usepackage[counterclockwise]{rotfloat}
\usepackage{epsf}
\newif\iffigs\global\figsfalse
\def\chkfig#1{
 \openin\epsffilein=#1
 \ifeof\epsffilein\figsfalse\else\figstrue\fi}

\def\bbox#1{\mathchoice
 {\mbox{\boldmath $\displaystyle#1$}}
 {\mbox{\boldmath $\textstyle   #1$}}
 {\mbox{\boldmath $\scriptstyle #1$}}
 {\mbox{\boldmath $\scriptstyle #1$}}}
\def\boldr{{\bbox{r}}}
\def\boldp{{\bbox{p}}}
\def\Re{\mathop{\rm Re}}
\def\Im{\mathop{\rm Im}}
\def\erf{\mathop{\rm erf}}
\def\arg{\mathop{\rm arg}}

\def\arccosh{\mathop{\rm arccosh}}
\def\rE{\mathop{\rm E}}
\def\rF{\mathop{\rm F}}
\def\Ai{\mathop{\rm Ai}}
\def\Gi{\mathop{\rm Gi}}
\def\mathatop#1#2{%
  \setbox0=\hbox{$#1$} \setbox1=\hbox{$#2$}
  \ifdim\wd0>\wd1 \setbox0=\hbox to\wd1{\hss\box0\hss}
  \else \setbox1=\hbox to\wd0{\hss\box1\hss} \fi
  \mathop{\vcenter{\offinterlineskip\box0\box1}}}
\def\lsim{\mathatop<\sim}
\def\gsim{\mathatop>\sim}

\def\dfrac#1#2{\displaystyle\frac#1#2}

\title{
\vspace{-25mm}\rightline{{\sf KUNS 1579}}\vspace{20mm}
Symmetry Breaking and Bifurcations
in the Periodic Orbit Theory:
I. Elliptic Billiard}
\author{%
Alexander G. {\sc Magner},$^{a,b,c}$
Sergey N. {\sc Fedotkin},$^{a,b}$ \\
Ken-ichiro {\sc Arita},$^d$
Toshiyuki {\sc Misu},$^e$
Kenichi {\sc Matsuyanagi},$^c$ \\
Thomas {\sc Schachner},$^b$
Matthias {\sc Brack}$^b$ \\[1ex]
\it
$^a$Institute for Nuclear Research, 252028 Prospekt Nauki 47,
Kiev--28, Ukraine \\
\it
$^b$Institute for Theoretical Physics, University of
Regensburg, \\
\it
D-93040 Regensburg, Germany \\
\it
$^c$Department of Physics, Graduate School of Science,
Kyoto University, \\
\it
Kitashirakawa, Kyoto 606-8502\\
\it
$^d$Department of Physics, Nagoya Institute of Technology,
Gokiso, Nagoya 466-8555 \\
\it
$^e$Cyclotron Radio-isotope Center, Tohoku University,
Sendai 980-8578}
\date{(May 27, 1999)}

\begin{document}
\maketitle

\begin{abstract}
We derive an analytical trace formula for the level density of the
two-dimensional elliptic billiard using an improved stationary phase
method.  The result is a continuous function of the deformation
parameter (eccentricity) through all bifurcation points of the short
diameter orbit and its repetitions, and possesses the correct limit of
the circular billiard at zero eccentricity.  Away from the circular
limit and the bifurcations, it reduces to the usual (extended)
Gutzwiller trace formula which for the leading-order families of
periodic orbits is identical to the result of Berry and Tabor.  We show
that the circular disk limit of the diameter-orbit contribution is also
reached through contributions from closed (periodic and non-periodic)
orbits of hyperbolic type with an even number of reflections from the
boundary.  We obtain the Maslov indices depending on deformation and
energy in terms of the phases of the complex error and Airy functions.
We find enhancement of the amplitudes near the common bifurcation
points of both short-diameter and hyperbolic orbits. The calculated
semiclassical level densities and shell energies are in good agreement
with the quantum mechanical ones.
\end{abstract}

\newpage
\tableofcontents

\newpage

\section{Introduction}

The periodic orbit theory (POT), developed by
Gutzwiller~\cite{gutzpr,gutz} for chaotic systems, by Balian and
Bloch~\cite{bablo} for cavities, and by Berry and Tabor~\cite{bt76,bt77}
for integrable systems, has proved to be an important semiclassical
tool not only for an approximate quantization, but also for the
description of gross-shell effects in finite fermion
systems~\cite{strusem,book}.  Gutzwiller's approach has been extended to
take into account continuous
symmetries~\cite{strusem,strupol,strutmag,creagh,mfimbrk,mfimb} and is
therefore applicable to systems with mixed classical dynamics,
including the integrable and hard-chaos limits.

An important role is played by the classical degeneracy of the
periodic orbits in systems with continuous spatial or dynamical
symmetries: the orbits are then not isolated in phase space (as it was
assumed in Gutzwiller's original trace formula, and as is the case in
chaotic systems), but occur in degenerate families with identical
actions.  The degree of degeneracy $\cal K$ is defined as the number
of independent parameters which are necessary to uniquely specify an
orbit within each family.  E.g., the orbit families with the highest
degeneracy in spherical systems with spatial $SO(3)$ symmetry have
${\cal K}=3$, corresponding to the three Euler angles that specify the
orientation of an orbit within the plane of motion and the orientation
of the plane itself; the orbit families in two-dimensional systems
with $U(1)$ rotational symmetry have ${\cal K}=1$; the isotropic
harmonic oscillator in 2 dimensions has $SU(2)$ symmetry and hence
orbit families with ${\cal K}=2$.  Orbits with different degeneracies
$\cal K$ may also occur in one and the same system, such as the
spherical cavity discussed by Balian and Bloch~\cite{bablo} where the
diameter orbit has ${\cal K}=2$ and all other orbits have ${\cal
K}=3$; the spheroidal cavity~\cite{msph} where ${\cal K}=2$, 1 or 0
occurs (the latter corresponding to isolated orbits); or the elliptic
billiard with ${\cal K}=1$ or 0, as discussed in the present paper.

However, problems arise for all these trace formulae in connection with
the breaking of a continuous symmetry and with the bifurcation of
stable periodic orbits when a continuous parameter (energy,
deformation, external field) is varied.  The reason is that at such
critical points the standard stationary phase approximation, used for
integrations in the derivation of the trace formula, breaks down and
leads to divergences and/or discontinuities of the amplitudes in the
trace formula.  This happens most frequently in mixed systems, but it
occurs also in integrable systems.  Typical examples are the
two-dimensional elliptic billiard and the three-dimensional spheroidal
cavity.  In the former, all repetitions of the short diameter orbits
undergo bifurcations at specific deformations, whereby new families of
hyperbolic orbits are created.  Similarly in the latter system, the
periodic orbits lying in the equatorial plane perpendicular to the
symmetry axis bifurcate also at specific deformations, whereby new
3-dimensional orbits appear~\cite{msph}.  In both systems, all
bifurcations and the limit to the spherical shape lead to divergent
amplitudes in the trace formulae, see
Refs.~\cite{strusem,mfimbrk,nish90,nish91,nish92,nish93,ak95,tdip,ak97,ask}.
Since for each family with a given value of $\cal K$, the extended
Gutzwiller trace formula~\cite{strusem,strupol,strutmag,creagh} has an
amplitude proportional to $\hbar^{-(1+{\cal K}/2)}$, it is evident that
the breaking of a continuous symmetry must be accompanied by a
discontinuous change of the amplitudes, which manifests itself in the
form of a singularity when one attempts to reach the unbroken symmetry
limit.  (An exceptional situation occurs in anisotropic harmonic
oscillators when changing from irrational to rational frequency ratios:
here the divergences of the different periodic orbit contributions have
been shown~\cite{brajain} to cancel identically, such that the trace
formulae --- which are quantum-mechanically exact here --- hold for
arbitrary frequency ratios, although their analytical form changes in
the different limits; see also Ref.~\cite{book}.)

Since symmetry breaking and orbit bifurcations occur in almost all
realistic physical systems, there is a definite need to overcome these
singularities.  The importance of bifurcation effects in connection
with the emergence of the `superdeformed' shell structure in atomic
nuclei was emphasized in Refs.~\cite{strusem,ak95,ak97,ask,ak98}. In
order to improve the POT in these critical situations, various methods
have been proposed.  Like in the treatment of continuous symmetries
considered in Refs.~\cite{strupol,strutmag,creagh,mfimbrk}, they
essentially consist in taking some integrals in the derivation of the
trace formula more exactly than by the standard stationary phase method
(SPM).

Berry and Tabor suggested in Ref.~\cite{bt76} a quite general method
to treat bifurcations in integrable systems.  Starting from the trace
integral for the level density in action-angle variables, they reduce
it to the Poisson-sum trace formula and do all trace integrations
except one by the SPM, extending the integration limits from $-\infty$
to $+\infty$.  At bifurcations, this leads to singularities in the
amplitudes when the stationary points are close to the limits of the
integration range.  According to Ref.~\cite{bt76}, in this case one
has to take the integral within the exact finite range.  The
integration range need not necessarily include the stationary points
(in the case of negative or complex stationary points), but the latter
are assumed to be close to the integration limits.  For integrable
systems, this idea was applied to the periodic-orbit families with the
highest degeneracies, for which one can exactly carry out the integrals
over the action angles, giving $2\pi$ for each degree of
freedom~\cite{bt77}.  This was the starting point of a uniform
approximation that was further developed by various
authors~\cite{richens,sie97ell,bremen}.

Another type of uniform approximation was initiated by Ozorio de
Almeida and Hannay~\cite{ozoriohan} (see also Ref.~\cite{ozorio}) and
further developed by Sieber and
Schomerus~\cite{sie96,sie97bif,schomsie} for various generic types of
bifurcations.  Writing the trace integral in a phase-space
representation, they expand the action around the bifurcation points
into so-called normal forms which usually can be integrated
analytically with finite results.  The correct asymptotic recovery of
the Gutzwiller amplitudes far from the bifurcation points can be
obtained by a suitable mapping transformation whereby the amplitude
function, together with the Jacobian of the mapping transformation, is
expanded up to an order consistent with that of the action in the
exponent of the integrand.  Near the bifurcation points, there is a
common contribution of all participating (real or complex, so-called
`ghost') orbits to the trace formula.

A similar technique, starting from the Berry-Tabor approach for
integrable systems and using a `pendulum mapping', was used by
Tomsovic, Grinberg and Ullmo~\cite{tomsovic,ullmo} to derive a generic
uniform approximation for the breaking of orbit families with a
one-dimensional degeneracy, corresponding to $U(1)$ symmetry, into
pairs of stable and unstable isolated orbits.  Finally, some analytical
uniform trace formulae for the breaking of the higher-dimensional
$SU(2)$ and $SO(3)$ symmetries in specific two- and three-dimensional
systems have been derived very recently~\cite{BMT}. Hereby the trace
integral was performed over the de Haar measure of the corresponding
symmetry groups, as in the derivation of the unperturbed trace
formulae for these continuous symmetries,~\cite{creagh} and the mapping
was done onto the forms of the action integrals obtained in
perturbation theory~\cite{Creaghpt,BCL}.

It should be mentioned that all the uniform approximations mentioned
above can be used only for one isolated critical point of symmetry
breaking or orbit bifurcation: they fail, in
particular,~\cite{sie96,sie97bif,schomsie,ullmo,BMT} when two critical
points are so close that the actions of the participating orbits at
these points differ by less than $\sim\hbar$.  To our knowledge, no
common uniform treatment of two nearby bifurcations (in the above
sense), or of a bifurcation near a symmetry-breaking point, has been
reported so far.

In this paper, we propose an approach to simultaneously overcome the
divergences due to symmetry breaking and any number of bifurcations in
the two-dimensional elliptic billiard and the three-dimensional
spheroidal cavity.  Although our framework is quite general, we limit
here its application to the elliptic billiard.  The three-dimensional
spheroidal cavity will be treated in a succeeding paper,~\cite{msph}
and the extension to non-integrable systems is planned for future
research.  We start from a phase-space trace
formula,~\cite{mfimbrk,mbfm} which after some transformations becomes
identical to that obtained from the mixed phase-space representation
of the Green function in Refs.~\cite{sie97bif,bruno}, as explained
there and further below in this paper (see \S4.3).  Analogous versions
of the phase-space trace formulae were suggested in
Refs.~\cite{bt77,creagh}.

In contrast to previous
investigations,~\cite{bt76,bt77,richens,sie97ell,bremen} we calculate
the integrals over angles, too, by the stationary phase method.  Note
that we also include orbits with lower degeneracies, such as the
isolated diameters in the elliptic billiard and the equatorial orbits
in the spheroidal cavity, hereby extending the method of
Ref.~\cite{bt76}.  Our main point is that the stationary-phase
integrals over both action and angle variables are calculated with
expansions of the phase and amplitudes like in the standard SPM, but
within {\em finite} intervals in all cases where it would lead to
divergences if one or both integration limits were taken to $\infty$
or $-\infty$. We will also discuss the role of non-periodic closed
orbits (see \S5.4).  For the Maslov indices, which for the bifurcating
orbits depend on the deformation and near the critical points also on
the energy, we follow the basic ideas of Maslov and
Fedoryuk~\cite{fed:jvmp,masl,fed:spm,masl:fed}.  We obtain separate
contributions to the trace formula from the bifurcating periodic
orbits, and we remove the singularity of the isolated long diameter
(i.e., the separatrix) near the circular shape of the elliptic
billiard in a simpler way than in Ref.~\cite{bremen}.

In this way we obtain an analytical trace formula for the elliptic
billiard which gives finite and continuous contributions at all
deformations, including the circular disk limit and all bifurcation
points of the short diameter orbit. Although its derivation and its
explicit form are quite different, our final trace formula is similar
to the uniform approximations mentioned above in the sense that it
connects smoothly to the standard (extended) Gutzwiller trace formulae
for the different orbit types for deformations sufficiently far away
from all critical points.

\section{Phase-Space Trace Formula in the Closed Orbit Theory}

\subsection{Semiclassical trace formula}
\label{traceformula}

The level density $g(\varepsilon)$ is obtained from the Green function
$G(\boldr',\boldr'';\varepsilon)$ by taking the imaginary part of its
trace:
\begin{eqnarray}
g(\varepsilon)
&=& -\frac{1}{\pi}\Im\int d\boldr'' \int d\boldr'\,
G(\boldr',\boldr''; \varepsilon) \delta(\boldr''-\boldr')\nonumber\\
&=& -\frac{1}{\pi}\Im\int d\boldr'' \int d\boldr' \int d\tilde{\boldp}\,
G(\boldr',\boldr''; \varepsilon)\,
\exp\left[-\frac{i}{\hbar}\tilde{\boldp}\cdot\left(
\boldr''-\boldr'\right)\right].
\label{trace}
\end{eqnarray}
Within the semiclassical Gutzwiller theory,~\cite{gutzpr,gutz} the
Green function $G(\boldr',\boldr''; \varepsilon)$ can be represented
in terms of the sum over all classical trajectories $\alpha$
connecting two spatial points $\boldr'$ and $\boldr''$ at fixed
energy $\varepsilon$.  Inserting it into (\ref{trace}), we obtain the
semiclassical level density
\begin{eqnarray}
g_{\rm scl}(\varepsilon) &=& \frac{2}{(2\pi \hbar)^{(3n+1)/2}}
\Im\sum_\alpha
\int d\boldr'' \int d\tilde{\boldp} \int d\boldr'
|{\cal J}(\boldp',t_\alpha; \boldr'',\varepsilon)|^{1/2}
\nonumber\\
&& \times\exp\Bigl\{\frac{i}{\hbar}\left[
S_\alpha(\boldr',\boldr'',\varepsilon)-\tilde{\boldp}\cdot(\boldr''-\boldr')
\right] - \frac{i\pi}{2}\mu_\alpha\Bigr\}.
\label{trace1}
\end{eqnarray}
Here $S_\alpha(\boldr',\boldr'',\varepsilon)=\int_{\boldr'}^{\boldr''}
d\boldr\cdot\boldp$ is the action along the trajectory $\alpha$, $n$ is
spatial dimension, and $\mu_\alpha$ is related to the number of {\em
conjugate points} (i.e., turning and caustics points along the
trajectory)~\cite{masl:fed}.  ${\cal J}_\alpha
(\boldp',t_\alpha;\boldr'',\varepsilon)$ is the Jacobian for the
transformation from initial momentum $\boldp'$ (at the point $\boldr'$)
and time interval $t_\alpha$ (for the classical motion along the
trajectory from initial to final point) to final coordinate $\boldr''$
and energy $\varepsilon$.

\subsection{Phase space variables}
\label{phasespace}

Integrating over $\boldr'$ in Eq.~(\ref{trace1}) along the direction
transverse to the trajectory $\alpha$ by the stationary phase method
(SPM), we are left with the integral over the component of $d\boldr'$
parallel to the trajectory, which gives just an energy conserving
delta function $\delta (\varepsilon-H(\boldr',\boldp'))$.  We hence
arrive at the phase-space trace formula~\cite{mbfm}
\begin{eqnarray}
g_{\rm scl}(\varepsilon)
&=& \frac{1}{(2\pi\hbar)^2}\Re\sum_\alpha
\int d\boldr'' \int d\boldp'\:
\delta(\varepsilon-H(\boldr',\boldp'))
\left|{\cal J}({\boldp}''_\perp,{\boldp}'_\perp)\right|^{1/2}
\nonumber \\
&& \times
\exp\Bigl\{\frac{i}{\hbar}\left[
S_\alpha(\boldp',\boldp'',t_\alpha)+(\boldp''-\boldp')\cdot\boldr''
\right] - i\nu_\alpha\Bigr\}.
\label{pstrace}
\end{eqnarray}
Here ${\cal J}({\boldp}''_\perp,{\boldp}'_\perp)$ is the Jacobian for
the transformation from initial to final momentum components
${\boldp}'_\perp$ and ${\boldp}''_\perp$, respectively, perpendicular
to the trajectory $\alpha$.  This Jacobian is equal to one of the
elements of the stability matrix (see, e.g., Ref.~\cite{book}).
$S_\alpha(\boldp',\boldp'',t_\alpha)$ is the action in the momentum
representation
\begin{equation}
S_\alpha(\boldp',\boldp'',t_\alpha)
= -\int_{\boldp'}^{\boldp''} d\boldp\cdot\boldr(\boldp),
\end{equation}
which is related to the usual action in coordinate space
\begin{equation}
S_\alpha(\boldr',\boldr'',\varepsilon)
= \int_{\boldr'}^{\boldr''} d\boldr\cdot\boldp(\boldr)
\label{actionr}
\end{equation}
by the Legendre transformation
\begin{equation}
S_\alpha(\boldr',\boldr'',\varepsilon)
- \boldp'\cdot(\boldr''-\boldr')
= S_\alpha(\boldp',\boldp'',t_\alpha)
+ (\boldp''-\boldp')\cdot\boldr''.
\label{legentrans}
\end{equation}

Note that the integrand in the phase-space trace formula
(\ref{pstrace}) (except for the exponent related to the phase part
proportional to $\boldr''$) is the semiclassical Green function in the
mixed representation which contains explicitly an energy-conserving
$\delta$-function in our case, unlike the form discussed in
Ref.~\cite{creagh}.  (Consequently, the momentum components are not
independent, which is important for the following application of the
stationary phase method; see more details in the next subsection and
in \S4.) Due to energy conservation, i.e., $H(\boldr',\boldp')\equiv
H(\boldr'',\boldp'')$, the trace formula (\ref{pstrace}) can be
rewritten in an alternative form where the integration variables are
changed from ($\boldr'',\boldp'$) to ($\boldr',\boldp''$).  The sum in
(\ref{pstrace}) runs over all isolated classical trajectories $\alpha$
with starting momentum $\boldp'$ and final point $\boldr''$ (or with
starting point $\boldr'$ and final momentum $\boldp''$ in the
alternative form), for a fixed time interval $t_\alpha$ of the
classical motion along $\alpha$.

\subsection{Periodic orbit theory}
\label{closedorbits}

The trajectories $\alpha$ in the phase space trace formula
(\ref{pstrace}) are not necessarily closed orbits in the usual
coordinate space.  But after separation of the extended Thomas-Fermi
part (corresponding to the `zero length orbits') and integration over
one of the momentum components exploiting the $\delta$-function, we
shall use further semiclassical approximations.  We first write the
stationary-phase conditions for the integration variables in
(\ref{pstrace}).  The stationary conditions for the momentum variable
$\boldp'$ are the closing condition for the trajectories $\alpha$ in
the usual coordinate space, $\boldr'=\boldr''$, and the Jacobian in
Eq.~(\ref{pstrace}) is unity due to the Liouville theorem of the
phase-space volume conservation, see Ref.~\cite{book}.  The additional
stationary-phase conditions for the integration over spatial variables
$\boldr''$ selects the periodic orbits, $\boldp'=\boldp''$, and we
obtain the POT and all known trace formulas including the Poisson-sum
trace formula~\cite{mbfm}.  We shall then integrate over components of
the phase-space variables exactly if we have identities for them.
Other integrations will be done by an improved stationary phase method
(ISPM). `Improved' here means that we carry out the integrations in
{\em finite} ranges, after expanding the exponent of the integrand
around the stationary point up to second order terms, and taking the
amplitude at the stationary point (or use a higher-order expansion of
amplitude and phase, if necessary).  All stationary points which appear
outside the physical region of the integration over the phase-space
variables are also taken into account, even if they are complex.  In
this way we get simple and continuous analytical solutions that stay
finite at all critical (bifurcation and symmetry-breaking) points.
Different from other uniform approximations mentioned in the
introduction, our results appear as explicit sums over separate
contributions that correspond to the periodic orbits in the asymptotic
regions away from the critical points.

\section{Classical Mechanics}
\label{clasmech}

\subsection{Elliptic billiard as integrable system}
\label{ellbill}

We consider an elliptic billiard with axes $a$ and $b$ (with $a\leq
b$) along the $y$ and $x$ coordinate axes, respectively, and ideally
reflecting walls.  This is an integrable system which can be separated
in the elliptic coordinates $(u,v)$ defined in terms of the cartesian
coordinates $(x,y)$ by
\begin{equation}
x=\zeta\cos u\sinh v, \qquad
y=\zeta\sin u\cosh v, \qquad
\zeta=\sqrt{b^2-a^2},
\label{ellcoord}
\end{equation}
with
\begin{equation}
-\frac{\pi}{2} \leq u \leq \frac{\pi}{2}, \qquad
0\leq v < v_b.
\label{ellcoordneq}
\end{equation}
Hereby $(x,y)=(\pm\zeta,0)$ are the foci of ellipses given by
$v=\mbox{const.}$, and $v=v_b$ is the elliptic boundary. It is
convenient to introduce the deformation parameter $\eta=b/a \geq 1$ and
to keep the area of the ellipse constant by setting $ab=R^2$, so that
one gets $b=R\sqrt{\eta}$ and $a=R/\!\sqrt{\eta}$.  The second constant
of the motion, besides the energy $\varepsilon$, is the product of the
angular momenta $l_{-}$ and $l_{+}$ with respect to the two foci.  For
the following, it is advantageous to use the single-valued quantity
$\sigma$ defined by
\begin{equation}
\sigma=1+\frac{l_{-}l_{+}}{2m\varepsilon\zeta^2}.
\label{sigma}
\end{equation}
There are two types of orbits, depending on the relative sign of
$l_{-}$ and $l_{+}$: {\em elliptic orbits} circulating around both
foci for $l_{-}l_{+}>0$ or $\sigma>1$, and librating {\em
hyperbolic orbits} for $l_{-}l_{+}<0$ or $\sigma<1$.  Their
names used here indicate that the former are limited to the area
between the elliptic boundary given by $v=v_b$ and a confocal elliptic
caustic given by $v=v_c$, whereas the latter are confined to the area
between the two branches of a hyperbolic caustic given by $u=\pm u_c$
and the elliptic boundary.  The critical values for the boundary and the
caustics are given by
\begin{equation}
v_b=\eta/\!\sqrt{\eta^2-1},\quad v_c=\arccosh(1/\!\sqrt{\sigma}),
\quad u_c=\arcsin(\sqrt{\sigma}).
\label{critic}
\end{equation}
In terms of the above quantities, the single-valued action integrals
$I_u$ and $I_v$ become
\begin{eqnarray}
I_u = \oint \! p_u du & = & \frac{p\,\zeta}{\pi}\! \int_{-u_c}^{u_c} \! du\,
 \sqrt{\sigma-\sin^2 u}, \nonumber \\
I_v = \oint \! p_v dv & = & \frac{p\,\zeta}{\pi}\! \int_{v_c}^{v_b} \! dv\,
 \sqrt{\cosh^2 v-\sigma},
\label{actionuv}
\end{eqnarray}
where $p=\sqrt{2m\varepsilon}=\hbar k$ is the constant classical
momentum of the particle.  Since the system is integrable, its
Hamiltonian depends only on the actions and not on the variables $u,v$,
i.e., $H(I_u,I_v,u,v) \equiv H(I_u,I_v)$.

\subsection{Periodic orbits}
\label{po}

As shown by Berry and Tabor,~\cite{bt76} the periodic orbits of an
integrable system are found by the condition that the angular
frequencies (for angle variables conjugate to the actions) have
rational ratios.  In the present case, these frequencies are given by
$\omega_u=\partial H/\partial I_u$, $\omega_v=\partial H/\partial I_v$,
so that the periodic orbits are characterized by pairs of positive
integers $(M_u,M_v)$
\begin{equation}
\frac{\omega_u}{\omega_v}\equiv \frac12\left[
1-\frac{\rF(\theta,\kappa)}{\rF(\frac{\pi}{2},\kappa)}
\right]=\frac{M_u}{M_v}, \qquad (M_u \geq 1,\; M_v \geq 2M_u),
\label{percond}
\end{equation}
where
\begin{equation}
\kappa = \sin u_c/\cosh v_c, \quad
\theta = \arcsin(\cosh v_c/\cosh v_b),
\label{kappathetac}
\end{equation}
and $\rF(\theta,x)$ is the elliptic integral of the first
kind~\cite{abramov}.  The greatest common divisor of $M_u$ and $M_v$
corresponds to the repetition number $M=1,2,3,\dots$ of a primitive
periodic orbit $(n_u,n_v)$:
\begin{equation}
(M_u,M_v)=(Mn_u,Mn_v)=M(n_u,n_v).
\label{primitive}
\end{equation}

\begin{figure}
\chkfig{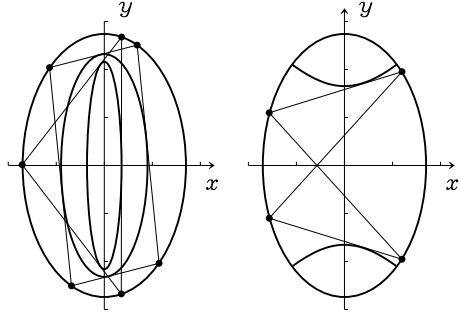}
\iffigs
\epsfxsize=.4\textwidth\centerline{\epsffile{fig01.ps}}
\else
\vspace{46mm}
\fi
\caption{\label{fig1}
Some classical periodic orbits in the elliptic billiard are shown
by thin solid lines. Left-hand side; elliptic
triangular $(1,3)$ and rhomboidal $(1,4)$ orbits:
Right-hand side; hyperbolic butterfly orbit $(1,4)$,
from Ref.~\protect\cite{mfimbrk}.}
\end{figure}

The solutions of Eq.~(\ref{percond}) for $\kappa$ and $\theta$ which
correspond to families of degenerate periodic orbits with ${\cal K}=1$
are, labeled accordingly for elliptic and hyperbolic orbits,
\begin{equation}
\left\{{{\kappa_e=\frac{1}{\sqrt{\sigma}}} \atop
        {\kappa_h=\sqrt{\sigma}}}\right\}, \quad
\left\{{{\theta_e=
         \arcsin\left(\sqrt{\sigma(1-1/\eta^2)}\right)} \atop
        {\theta_h=
         \arcsin\left(\sqrt{1-1/\eta^2}\right)}}\right\}.
\label{kappatheta}
\end{equation}
Figure \ref{fig1} shows the shortest periodic orbits of each kind.
The degeneracy parameter ${\cal K}$ was defined as the number of parameters
that specify the orbits within a family with a common action. Due to
the separation of variables in elliptic coordinates (\ref{ellcoord}) we
have two single-valued action integrals $I_u$ and $I_v$
(\ref{actionuv}).  They are related through the energy conserving
equation $\varepsilon=H(I_u,I_v)$, and can be written in terms of one
parameter of the family $\sigma$ (or $l_{-}l_{+}$), i.e., we have
${\cal K}=1$ (see
Refs.~\cite{strusem,strupol,strutmag,mfimbrk,landlifclm} for more
details.)

\subsection{Energy surface}
\label{energy}

\begin{figure}
\chkfig{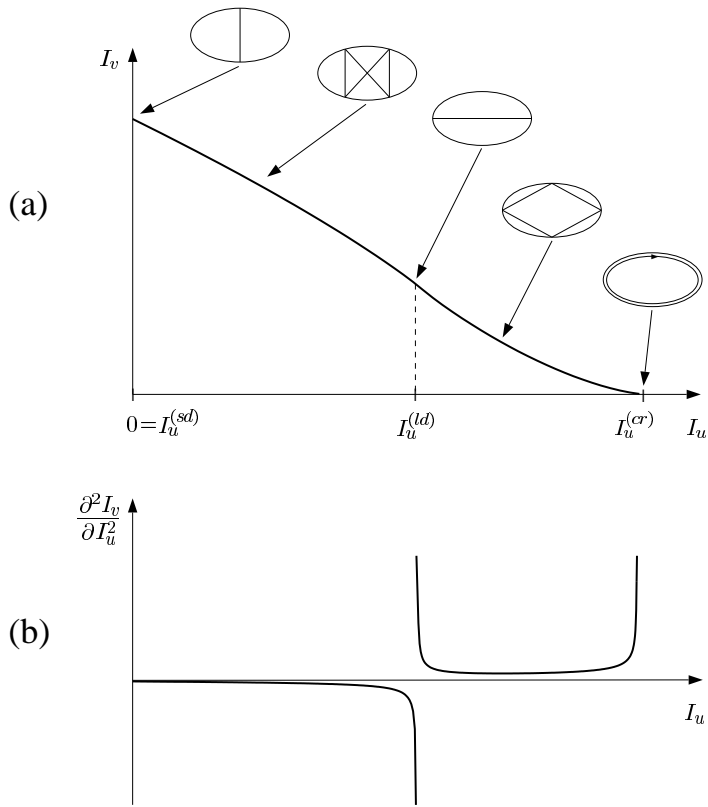}
\iffigs
\epsfxsize=.6\textwidth\centerline{\epsffile{fig02.ps}}
\else
\vspace{113mm}
\fi
\caption{\label{fig2}
Energy surface $I_v(I_u)$ and curvature $\partial^2
I_v/\partial I_u^2$ are drawn in the upper and lower panels,
respectively, from Ref.~\cite{tdip}.}
\end{figure}

For the energy surface $\varepsilon=H(I_u,I_v)$ one can get from
Eqs.~(\ref{actionuv}) the parametric equations (\ref{ensurfe}) for the
elliptic orbits and (\ref{ensurfh}) for the hyperbolic
orbits~\cite{tdip}.  The energy curve (\ref{ensurfe}) or
(\ref{ensurfh}) can also be considered through the single-valued
parameter $\sigma$ or double-valued $\kappa$ defined within the same
range $0 \leq \kappa
\leq 1$ for both kinds of orbits.  The solutions $\sigma$ found from
the periodic orbit equations (\ref{percond}) for elliptic orbits
satisfy the inequality $\sigma > 1$ in the elliptic part
(\ref{ensurfe}) of the energy curve.  On the other hand, $\sigma < 1$
for the hyperbolic part (see Fig.~\ref{fig2}a).  The two regions are
separated by the separatrix point $\sigma_s=1$, corresponding to the
long diameter orbit, where the value of the action $I_u=I_u^{(s)}$ is
given by
\begin{equation}
I_u^{(s)}=2p\zeta/\pi.  \qquad (\sigma_s=1)
\label{iusep}
\end{equation}
Thus, each phase space torus is split into two regions by the
separatrix: a hyperbolic and an elliptic region.  In the hyperbolic part
($0\leq\sigma<1$), the action variable $I_u$ changes from $0$ to the
separatrix value $I_u^{(s)}$.  In the elliptic part ($1<\sigma\leq
\sigma_{\rm cr})$, $I_u$ changes from the separatrix value to the
maximum value $I_u^{\rm(cr)}$ that corresponds to a `creeping' (or
`whispering gallery') orbit and is given by
\begin{eqnarray}
I_u^{\rm(cr)}
& = & \frac{2pR\sqrt{\eta}}{\pi}
\,\rE\!\left(\frac{\pi}{2},\frac{1}{\sqrt{\sigma_{\rm cr}}}\right)
= \frac{2pR\sqrt{\eta}}{\pi}
\,\rE\!\left(\frac{\pi}{2},\frac{\sqrt{\eta^2-1}}{\eta}\right)\!,
\nonumber \\
\sigma_{\rm cr} & = & \cosh^2 v_b=\eta^2/(\eta^2-1).
\label{iucr}
\end{eqnarray}

The short diameter (1,2) and its repetitions $M$(1,2) correspond to the
end point of the hyperbolic region at $\sigma=0$ ($\kappa=0$), which is
isolated in phase space $\{\Theta_u,I_u\}$. Eq.~(\ref{percond}) for the
periodic orbits at this $\sigma$ can be solved analytically with
respect to $\theta$.  Identifying the root $\theta(\eta,n_u/n_v)$ with
its definition (\ref{kappatheta}) for hyperbolic orbits we realize that
all short diameters $M$(1,2) bifurcate at the deformations,
\begin{equation}
\eta_{\rm bif}(M,n)=\frac{1}{\sin(\pi n_u/n_v)}=
\frac{1}{\cos(n\pi/2M)},\quad (n=1,2,3,\cdots,M-1)
\label{etamin}
\end{equation}
and at each bifurcation a new family of hyperbolic orbits $M(n_u,n_v)$
with $Mn_v$ reflection points is `born'. The second equation presents
the same bifurcation points and shows explicitly that the bifurcation
deformations $\eta_{\rm bif}$ are also identical to the corresponding
divergences of the Gutzwiller amplitudes for short diameters, see
Eq.~(6.47) of Ref.~\cite{book}. Each of the emerging hyperbolic orbits
$M_1(M-n,2M)$ with $M_1$ repetitions and $n$ from Eq.~(\ref{etamin})
coincides exactly with the corresponding short diameter $M_1M$(1,2)
repeated $M_1M$ times at the deformation $\eta_{\rm bif}$. For
instance, for the triply repeated short diameter 3(1,2) ($M_1=1,M=3$)
there are two bifurcation points at the deformations
$\eta_{bif}=2/\sqrt{3}$ and $2$ where the primitive hyperbolic orbits
(2,6) ($n=1$) and (1,6) ($n=2$), respectively, are born (see these
orbits in Fig.~3.6 and discussion nearby in Ref.~\cite{tdip}, also
Ref.~\cite{nish90} and Fig.~1a there). However, the short
diameters are {\em isolated} in the phase space of action-angle
variables $\{\Theta_u$,$I_u\}$. They emerge as terms of the periodic
orbit sum which are additional to the families of hyperbolic tori (see
a more detailed discussion below). The contribution of the primitive
short diameter 1(1,2) can be calculated by the original
Gutzwiller trace formula, except near the circular
shape~\cite{book,tdip}.  This formula will be improved near all
bifurcation points (\ref{etamin}) and the circular shape in \S5.2.

The long diameter orbits $M$(1,2) are also characterized by 2$M$
reflection points and correspond to a specific {\em isolated} point in
$\{\Theta_u,I_u\}$ space. They are related to the separatrix value
$\sigma=1$ ($\kappa=1$). Again, their amplitudes can be calculated with
the standard Gutzwiller trace formula for isolated orbits, with the
same exception near the symmetry-breaking point of the circular
shape~\cite{book,tdip} (see \S5.3 for the improved solution in terms of
Airy functions near this point).

The limit of the circular disk ($\eta=1$) may in some sense also be
considered as a (one-sided) bifurcation point: here the family of
diameter orbits (with ${\cal K}=1$) break into two isolated diameters
with ${\cal K}=0$ and complicated hyperbolic orbit families (${\cal
K}=1$) with $n_u \rightarrow \infty$, $n_v \rightarrow \infty$, and
$n_u:n_v \rightarrow 1:2$, when the deformation ($\eta>1$) is turned
on. Inversely, the long and short diameters and hyperbolic orbits which
have ${\cal K}=0$ and 1 in the ellipse, respectively, merge into the
families of diameter orbits with ${\cal K}=1$ as $\eta\rightarrow 1$.
The discontinuous change of $\cal K$ at $\eta=1$ is accompanied by a
divergence of the diametric amplitudes in the standard SPM. This is the
symmetry breaking problem discussed in the introduction and below in
\S5.2 and \S5.3.

Figure \ref{fig2}(a) shows the energy surface in action space, in the
form of the curve $I_v=I_v(\varepsilon,I_u)$ at fixed energy
$\varepsilon$. Specific primitive orbits (with $M=1$) are illustrated,
with the arrows pointing to the corresponding stationary points
$I_u^*$: the short diameter (at $I_u^*=0$ or $\sigma=0$, with
$\Theta_u^*=0,\pi$), the `butterfly' (or `bow-tie') orbit, the long
diameter (at $I_u^*=I_u^{(s)}$, with $\sigma=1$ and
$\Theta_u^*=\pm\pi/2$), the rhomboidal orbits with 4 reflections, and
the `creeping' orbit (at $I_u^*=I_u^{\rm(cr)}$) as the limit of a
`whispering-gallery' mode with number of reflections $n_v=\infty$ and
winding number $n_u=1$.  The limits to the separatrix correspond to
infinite values of $n_v$ and $n_u$ for hyperbolic or elliptic orbits
with the ratio $n_u/n_v$ going to $1/2$ from either side (see also
Ref.~\cite{nish90}). We use the same notation for both short and long
diameters in terms of the integers $n_u,n_v$ and $M$ like for the
elliptic and hyperbolic one-parametric families, specifying them also
by the stationary points in the phase space variables $\sigma$ (or
$I_u$) for all orbits and $\Theta_u$ for the isolated ones if
necessary.

\subsection{Curvature}
\label{curv}

A key quantity in the semiclassical theory in terms of the action-angle
variables is the curvature $K$ of the energy surface
\begin{equation}
K = \frac{\partial^2 I_v}{\partial I_u^2}
  = \Big(
    \frac{\partial^2 I_v}{\partial \sigma^2}
    + \frac{\omega_u}{\omega_v}
    \frac{\partial^2 I_u}{\partial\sigma^2}
    \Big)\Big/
    \Big(\frac{\partial I_u}{\partial \sigma}\Big)^2.
\label{curvature}
\end{equation}
The partial derivatives appearing on the right-hand side above are
given in Appendix A. Figure \ref{fig2}(b) shows $K$ versus $I_u$. In
the limit $\sigma \to 0$ one finds the curvature for the twice repeated
short diameters considered as primitive orbits~\cite{tdip}.  For
our definition of the (non-repeated) primitive orbits, one has the curvature
$K_s$ larger by a factor 2, i.e.,
\begin{equation}
K_s = -\frac{1}{\pi pR\eta^{3/2}}
\label{curvshortd}
\end{equation}
which is finite and negative for all deformations. $K$ stays negative
for the entire hyperbolic part $0\leq\sigma<1$ of the curve, whereas it
is positive for the elliptic part $1<\sigma<\sigma_{\rm cr}$. At the
critical points $\sigma=1$ (separatrix) and at $\sigma_{\rm cr}$
(creeping point), the curvature diverges. It tends to $-\infty$ as one
approaches the separatrix from the hyperbolic side, and to $+\infty$
from the elliptic side. For $\sigma\rightarrow\sigma_{\rm cr}$ it also
tends to $+\infty$.

\section{Phase Space Trace Formula in Action-Angle Variables}

\subsection{Action-angle variables}
\label{actang}

We now transform the phase space trace formula (\ref{pstrace}) from
the usual phase space variables $(\boldr,\boldp)$ to the angle-action
variables $(\bbox{\Theta},\bbox{I})$.  The latter are useful for
integrable systems because the Hamiltonian $H$ does not depend on the
angle variables $\bbox{\Theta}$, i.e., $H=H(\bbox{I})$.  For elliptic
billiard one has from (\ref{pstrace})
\begin{eqnarray}
g_{\rm scl}(\varepsilon)\
&=& \frac{1}{(2\pi\hbar)^2}\Re\sum_\alpha
\int d\Theta_u''\int d\Theta_v''\int dI_u'\int dI_v'\,
\delta(\varepsilon-H(I_u',I_v')) \nonumber\\
&& \times \exp\left\{\frac{i}{\hbar}\left[
S_\alpha(\bbox{I}',\bbox{I}'',t_\alpha)+
(\bbox{I}''-\bbox{I}')\cdot\bbox{\Theta}''\right]
-i\nu_\alpha\right\},
\label{pstraceactang}
\end{eqnarray}
where $\bbox{\Theta}=\{\Theta_u,\Theta_v\}$ are the angles and
$\bbox{I}=\{I_u,I_v\}$ the actions for the elliptic billiard defined
in the previous section.  For simplicity we omitted here and below the
Jacobian pre-exponential factor of Eq.~(\ref{pstrace}) because this
Jacobian taken at the stationary points is always unity when we apply
the improved stationary phase method for the calculation of the
integral over phase space variables, as noted above.

\subsection{Stationary phase method and classical degeneracy}
\label{spmdeg}

As noted in the introduction, we emphasize that {\em even for
integrable systems} the trace integral (\ref{pstraceactang}) is more
general than the Poisson-sum trace formula which is the starting point
of Refs.~\cite{bt76,bt77} for the semiclassical derivations.  These
two trace formulae become identical when we assume that the phase of
the exponent also does not depend on the angle variables
$\bbox{\Theta}$, like the Hamiltonian.  Then, the integral over angles
in (\ref{pstraceactang}) simply gives $(2\pi)^n$ where $n$ is the
spatial dimension ($n=2$ for the elliptic billiard), see
Ref.~\cite{bt77}.  In this case the stationary condition for all angle
variables are identities in the $2\pi$ interval.  This is true for the
contribution of the most degenerate classical orbits like elliptic and
hyperbolic orbits with ${\cal K}=1$ in the elliptic billiard.  For the
case of orbits with smaller degeneracy like the isolated diameters
(${\cal K}=0$) in the elliptic billiard, the exponent phase is a
strongly dependent function of some angles with definite discrete
stationary points.  We therefore need to integrate over such angles by
the standard or improved SPM.  Other examples are the equatorial orbits
(${\cal K}=1$) and diameters along the symmetry axis (separatrix with
${\cal K}=0)$ in the spheroidal cavity ($n=3$), the degeneracy
parameters of which are smaller than the largest possible value ${\cal
K}={\cal K}_{\rm max}=2$ for the elliptic and hyperbolic orbits in the
meridian plane, or for 3-dimensional orbits.  We have a similar
situation also for the diameters with ${\cal K}=2$ in the spherical
cavity (${\cal K}_{\rm max}=3$), orbits along the symmetry axis for
axially-symmetric cavities, and so on.  Thus, the stationary conditions
with respect to the angle variables for orbits with smaller
degeneracies are not identities. Moreover, the stationary points in the
cases mentioned above occupy subspaces of the phase space which are
isolated in the rational tori that lead to separate contributions to
the trace formula, except for the most degenerate orbit families, as we
shall see below for the case of the elliptic billiard.

\subsection{Stationary phase conditions}
\label{statcondactang}

We first take the integral over $I_v'$ in Eq.~(\ref{pstraceactang})
exactly.  Due to the energy conserving $\delta$-function, we are left
with the integrals over angles $\Theta_u''$, $\Theta_v''$ and action
$I_u'$:
\begin{eqnarray}
g_{\rm scl}(\varepsilon)
&=& \frac{1}{(2\pi\hbar)^2}\Re\sum_\alpha
\int d\Theta_u''\int d\Theta_v''\int dI_u'\,
\frac{1}{|\omega_v'|} \nonumber \\
&& \times\exp\left[\frac{i}{\hbar}\left(
S_\alpha(\bbox{I}',\bbox{I}'',t_\alpha)
+(\bbox{I}''-\bbox{I}')\cdot\bbox{\Theta}''\right)
-i\nu_\alpha\right],
\label{pstraceactang1}
\end{eqnarray}
\begin{equation}
S_\alpha(\bbox{I}',\bbox{I}'',t_\alpha)
= -\int_{\bbox{I}'}^{\bbox{I}''} d\bbox{I}\cdot\bbox{\Theta}(\bbox{I}).
\label{actionactvar}
\end{equation}

We first write down the stationary phase equation for $I_u$:
\begin{equation}
\left(\frac{\partial S_\alpha(\bbox{I}',\bbox{I}'',t_\alpha)}{
\partial I_u'}\right)^* - \Theta_u''
\equiv \Theta_u'-\Theta_u''=2\pi M_u,
\label{statcondi}
\end{equation}
where $M_u$ is an integer.  The star means that we take the quantities
at the stationary point $I_u'=I_u^*$.  We now use the
Legendre transformation (\ref{legentrans}), which reads
\begin{equation}
S_\alpha(\bbox{I}',\bbox{I}'',t_\alpha)
+(\bbox{I}''-\bbox{I}')\cdot\bbox{\Theta}''
= S_\alpha(\bbox{\Theta}'',\bbox{\Theta}',\varepsilon)
-\bbox{I}'\cdot(\bbox{\Theta}''-\bbox{\Theta}'),
\label{phasecondti}
\end{equation}
\[
S_\alpha(\bbox{\Theta}',\bbox{\Theta}'',\varepsilon)
= \int_{\bbox{\Theta}'}^{\bbox{\Theta}''}
d\bbox{\Theta}\cdot\bbox{I}(\bbox{\Theta}).
\]
With the use of this transformation, the stationary phase conditions
for angles $\Theta_u$ and $\Theta_v$ are written as
\begin{equation}
\left(\frac{\partial
S_\alpha(\bbox{\Theta}',\bbox{\Theta}'',\varepsilon)}{
\partial\bbox{\Theta}''}
+\frac{\partial
S_\alpha(\bbox{\Theta}',\bbox{\Theta}'',\varepsilon)}{
\partial\bbox{\Theta}'}\right)^*
\equiv \bbox{I}''-\bbox{I}'=0.
\label{statcondt}
\end{equation}

For the following derivations we have to decide which stationary phase
conditions from Eqs.~(\ref{statcondi}) and (\ref{statcondt}) are
identities for the finite volume of the phase-space tori and which are
equations for the isolated stationary points. For doing this, we have
to calculate separately the contributions from the most degenerate
(elliptic and hyperbolic) families $({\cal K}=1)$ to the improved trace
formula and those from diameters in the elliptic billiard. These two
contributions are different with respect to the above mentioned
decision concerning the integration over the angles $\bbox{\Theta}$.
After the integration over one of the angle variables, say $\Theta_v$,
corresponding to the identity in the stationary phase conditions
(\ref{statcondt}) due to an invariance of the action along the periodic
orbit in Eq.~(\ref{pstraceactang1}), one gets Eq.~(7) of
Ref.~\cite{sie97bif} derived earlier by Bruno~\cite{bruno}.  So, we get
the result of Refs.~\cite{sie97bif,bruno} within periodic orbit
theory. Our phase-space trace formula (\ref{pstrace}) is more general
because it can be applied for more exact calculations of the level
density, without use of the stationary phase conditions like
Eqs.(\ref{statcondt}), in terms of closed (periodic and non-periodic)
orbits.

Note that we have separate contributions coming from each kind of
families and isolated orbits {\em even near the bifurcation points}
(\ref{etamin}) where we have the end point. Taking the deformation at
a small distance from $\eta_{\rm bif}$, we are left with two separate
close stationary points and then use the Maslov-Fedoryuk
theory~\cite{fed:jvmp,masl,fed:spm,masl:fed} like for caustic and
turning points. Finally, after the integration by the improved
stationary phase method, we look at the limit $\eta \rightarrow
\eta_{\rm bif}$ to the bifurcation point. In particular, this idea of
Maslov and Fedoryuk was applied in Appendix B for the calculation
of the contribution of the long diameter at the separatrix.

\section{Trace Formulas for the Elliptic Billiard}

\subsection{Elliptic and hyperbolic orbit families $({\cal K}=n-1=1)$}

Each family of elliptic or hyperbolic orbits with a common action
occupies a two-dimensional finite area in the elliptic billiard. In
this case, the stationary conditions (\ref{statcondt}) for the
integration over the angle variables $\Theta_u$ and $\Theta_v$ become
identities, since the integrand does not depend on the angle variables,
and we have the conservation of the action variable $I_u'=I_u''=I_u$
fulfilled identically along each classical trajectory $\alpha$. Taking
the integrals over $\bbox{\Theta}$ gives a factor $(2\pi)^2$, and we
are left with the Poisson-sum trace formula like in
Refs.~\cite{bt76,bt77}:
\begin{eqnarray}
g_{\rm scl}(\varepsilon)
&=& \frac{1}{\hbar^2}\Re\sum_{\bbox{M}}
\int d\bbox{I}\:\delta(\varepsilon-H(\bbox{I}))
\exp\left[\frac{2\pi i}{\hbar} \bbox{M}\cdot\bbox{I}-i\nu_{\bbox{M}}
\right] \nonumber \\
&=& \frac{1}{\hbar^2}\Re\sum_{\bbox{M}}
\int dI_u\:\frac{1}{|\omega_v|}
\exp\left[\frac{2\pi i}{\hbar} \bbox{M}\cdot\bbox{I}-i\nu_{\bbox{M}}
\right].
\label{poissonsum}
\end{eqnarray}
Here $\bbox{M}=(M_u,M_v)$ are integers which correspond to those in
Eq.~(\ref{primitive}). Next we transform the integration variable in
the last expression of Eq.~(\ref{poissonsum}) from $I_u$ to $\sigma$
defined by (\ref{sigma}). Thus, the level density component $\delta
g_{{\rm scl},1}$ related to the one-parameter families can be written
as a sum of contributions from the hyperbolic ($\delta g_{{\rm
scl},1}^{(h)}(\varepsilon)$) and the elliptic ($\delta g_{{\rm
scl},1}^{(e)}(\varepsilon)$) parts of the tori. Their sum is
\begin{equation}
\delta g_{{\rm scl},1}(\varepsilon)
= \frac{1}{\pi\varepsilon_0 pR^2}\Re\!\sum_{\bbox{M}}
\frac{1}{n_v}\!\int_0^{\sigma_{\rm cr}}\!\! d\sigma
L_{\bbox{M}}\frac{\partial I_u}{\partial \sigma}
\exp\!\!\left[\frac{2\pi i}{\hbar}\bbox{M}\!\cdot\!\bbox{I}(\sigma)
-i\nu_{\bbox{M}}\right]\!,
\label{poissonsums}
\end{equation}
where $\varepsilon_0=\hbar^2/(2mR^2)$, $\bbox{I}(\sigma)$ are the
actions defined by Eqs.~(\ref{actionuv}), $L_{\bbox{M}}$ are the
`lengths' of the primitive orbits with $M=1$ given by
\begin{eqnarray}
L_{\bbox{M}}
&=& \frac{2\pi n_v p}{m\omega_v} \nonumber\\
&=& 2n_v b\,\sin{\theta}\!\left[\rE(\theta,\kappa)
-\frac{\rF(\theta,\kappa)}{\rF(\frac{\pi}{2},\kappa)}\,
\rE({\textstyle\frac{\pi}{2}},\kappa)+\cot\theta
\sqrt{1-\kappa^2\sin^2\theta}\right],
\label{length}
\end{eqnarray}
and $\theta(\sigma)$ and $\kappa(\sigma)$ are defined by Eq.\
(\ref{kappatheta}). The `lengths' become the true lengths of the
corresponding periodic orbits when they are taken at $\sigma$ equal to
the real positive roots of Eq.~(\ref{percond}) inside the integration
range. For other values of $\sigma$, the `lengths' are nothing else
than the functions (\ref{length}) introduced in place of $\omega_v$ for
convenience.  The integration range from the bifurcation point
$\sigma=0$ to the separatrix $\sigma_s=1$ covers the contributions of
all hyperbolic orbits. The remaining part of Eq.~(\ref{poissonsums})
from $\sigma=1$ to the creeping value $\sigma_{\rm cr}$ gives the
contributions from the elliptic tori.

As we shall see below, the choice of $\sigma$ as the integration
variable significantly improves the precision of the SPM. We hence
apply the stationary condition (\ref{statcondi}) for the phase in the
integrands of Eq.~(\ref{poissonsums}) with respect to $\sigma$ rather
than to $I_u$. With Eqs.~(\ref{kappatheta}), this condition becomes
identical to Eq.~(\ref{percond}) and determines the stationary phase
point $\sigma'=\sigma''=\sigma^*$ related to $I_u'=I_u''=I_u^*$.  We
used here the conservation of $\sigma$ (or the additional integral of
motion $l_{+}l_{-}$) along the periodic orbit. We now expand the phase
up to second order,
\begin{equation}
S_\alpha(\bbox{I}',\bbox{I}'',t_\alpha)
+ (\bbox{I}''-\bbox{I}')\cdot\bbox{\Theta}''
= 2\pi \bbox{M}\cdot\bbox{I}
= S_\beta(\varepsilon)+\frac12 J_\beta^\parallel(\sigma-\sigma^*)^2,
\label{phaseexp}
\end{equation}
where $S_\beta$ is the action along the periodic orbit $\beta$
determined by Eq.~(\ref{percond}),
\begin{equation}
S_\beta(\varepsilon) = 2\pi M(n_u I_u(\sigma^*)+n_v I_v(\sigma^*)),
\label{actionpo}
\end{equation}
and $J_\beta^\parallel$ is the Jacobian stability factor with
respect to $\sigma$ along the energy surface:
\begin{equation}
J_\beta^\parallel
=\left(\frac{\partial^2 S}{\partial \sigma^2}
\right)_{\sigma=\sigma^*,\beta}
=2\pi M\left(n_u\frac{\partial^2 I_u}{\partial \sigma^2}
            +n_v\frac{\partial^2 I_v}{\partial \sigma^2}
\right)_{\sigma=\sigma^*,\beta}.
\label{jacobpar0}
\end{equation}
It is related to the curvature $K_\beta$ (\ref{curvature}) of the
energy surface by
\begin{equation}
J_{\beta}^\parallel
= 2\pi M n_v K_\beta
\left(\frac{\partial I_u}{\partial\sigma}\right)^2_{\sigma=\sigma^*,\beta}
= 2\pi M n_v \epsilon \left|K_\beta\right|
\left(\frac{\partial I_u}{\partial\sigma}\right)^2_{\sigma=\sigma^*,\beta},
\label{jacobpar}
\end{equation}
where $\epsilon=+1$ for elliptic orbits and $\epsilon=-1$ for
hyperbolic orbits. We substitute now the expansion (\ref{phaseexp}) and
take the pre-exponential factor off the integral in
Eq.~(\ref{poissonsums}). For the sake of simplicity, we only consider
the lowest order in the expansion of the phase and the pre-exponential
factor in Eq.~(\ref{poissonsums}) in the variable $\sigma$, although
higher-order expansions can in principle be used to improve the
precision of the SPM. Thus, we are left with the integral from
$\sigma=0$ to $1$ for the hyperbolic orbits, and from $\sigma=1$ to
$\sigma_{\rm cr}$ for the elliptic orbits.

When the stationary point $\sigma^*$ is far from the limits of these
intervals, one can extend the integration range from $-\infty$ to
$\infty$ and get the result of the standard POT~\cite{bt76}.  Near the
bifurcation points (\ref{etamin}) of the short diameter orbit (where
the hyperbolic orbit families appear), however, the stationary point
$\sigma^*$ is close to zero. In this case we cannot extend the lower
limit to $-\infty$ but have to take the integral exactly from
$\sigma=0$. On the other hand, when the stationary point $\sigma^*$
approaches the integration limits $\sigma_s$ (\ref{iusep}) or
$\sigma_{\rm cr}$ (\ref{iucr}), hyperbolic or elliptic orbits with an
increasing number of corners $n_v$ appear.  In these cases, too, we
cannot extend the integration limits to $\pm\infty$.  Taking the
integral over $\sigma$ within the {\em finite limits}, we obtain a
trace formula in terms of complex Fresnel functions or generalized
error functions. The contributions of the one-parameter orbit families
$\delta g_{{\rm scl},1}(\varepsilon)$ are then given in the form
\begin{equation}
\delta g_{{\rm scl},1}(\varepsilon) = \Re\sum_\beta
{\cal A}_\beta^{(1)}(\varepsilon)
\exp\left[ikL_\beta-i\nu_\beta^{\rm(tot)}\right].
\label{deltag1}
\end{equation}
Here, the sum is taken over both elliptic and hyperbolic orbit
families, $k=\sqrt{2m\varepsilon}/\hbar$. The amplitude
${\cal A}_\beta^{(1)}$ of the orbit family $\beta$ is given by
\begin{equation}
{\cal A}_\beta^{(1)}=\frac{L_\beta}{
2\varepsilon_0\pi kR^2\sqrt{-\epsilon i M^3 n_v^3
\left|\hbar K_\beta\right|}}
\erf\left({\cal Z}_{\beta,1}^\parallel,{\cal Z}_{\beta,2}^\parallel\right);
\label{amp1}
\end{equation}
$L_\beta$ is the `length' of the orbit family (\ref{length}) 
corresponding to the stationary point $\sigma^*$ ($M=1$). We have
introduced here the generalized error function $\erf(z_1,z_2)$:
\begin{equation}
\erf(z_1,z_2)=\frac{2}{\sqrt{\pi}}\int_{z_1}^{z_2} dz e^{-z^2}
=\erf(z_2)-\erf(z_1),
\label{errorf}
\end{equation}
$\erf(z)$ being the standard error function~\cite{abramov} with
(complex) argument $z$. The complex quantities ${\cal
Z}_{\beta,1}^\parallel$ and ${\cal Z}_{\beta,2}^\parallel$ in
(\ref{amp1}) are given in terms of the Jacobian $J_{\beta}^\parallel$
(\ref{jacobpar0}) and the stationary points $\sigma^*$:
\begin{equation}
{\cal Z}_{\beta,1}^\parallel
= \sqrt{\frac{\epsilon i |J_{\beta}^\parallel|}{2\hbar}}
\left(\sigma_{\rm min}^{(\epsilon)}-\sigma^*\right), \qquad
{\cal Z}_{\beta,2}^\parallel
= \sqrt{\frac{\epsilon i |J_{\beta}^\parallel|}{2\hbar}}
\left(\sigma_{\rm max}^{(\epsilon)}-\sigma^*\right),
\label{argerrorpar}
\end{equation}
where $\sigma_{\rm min}^{(\epsilon)}$ and $\sigma_{\rm
max}^{(\epsilon)}$ are related to the integration limits by
\begin{equation}
\sigma_{\rm min}^{(\epsilon)} = \left\{
 \begin{array}{c@{,\quad}l}
  1 & \epsilon=1 \\
  0 & \epsilon=-1
 \end{array}\right\}, \qquad
\sigma_{\rm max}^{(\epsilon)} = \left\{
 \begin{array}{c@{,\quad}l}
 \sigma_{\rm cr} & \epsilon=1 \\
 1               & \epsilon=-1
 \end{array}\right\}.
\label{sigmaminmax}
\end{equation}
The phases $\nu_\beta^{\rm(tot)}$ in (\ref{deltag1}) are related to
the Maslov indices. They have a constant part $\nu_\beta$ which is
independent of deformation $\eta$ and energy $\varepsilon$.  At
deformations which are far enough from bifurcation points, such that
the stationary points are far enough from the integration limits, we
can determine this asymptotic part $\nu_\beta$ by transforming the
error functions to Fresnel functions~\cite{abramov} with real limits
and extending the integration limits to $\pm \infty$.  We hereby arrive
at the amplitude ${\cal A}_\beta^{(1)}$ of the standard
POT~\cite{bt76,mfimbrk,frisk}
\begin{equation}
{\cal A}_\beta^{(1)}=\frac{L_\beta}{
\varepsilon_0 \pi kR^2 \sqrt{-\epsilon i M^3 n_v^3
\left|\hbar K_\beta\right|}},
\label{bt}
\end{equation}
and $\nu_\beta$ is determined by the number of turning and caustic
points as in the theory of Maslov and
Fedoryuk~\cite{fed:jvmp,masl,fed:spm,masl:fed}.  In terms of the numbers
$n_v$ and $n_u$ and the repetition number $M$, it is given by
\begin{eqnarray}
 \nu_\beta &=& \dfrac{3\pi}{2}n_vM \hspace{62pt}
 \mbox{for $\epsilon=+1$}, \nonumber\\
 \nu_\beta &=& \dfrac{\pi}{2}(2n_u+2n_v)M \qquad
 \mbox{for $\epsilon=-1$}.
\label{maslovphase}
\end{eqnarray}
>From Eqs.~(\ref{deltag1}), (\ref{amp1}),
and (\ref{maslovphase}) we determine an extra contribution to the total
phase $\nu_\beta^{\rm(tot)}$
\begin{equation}
\nu_\beta^{\rm(tot)} = \nu_\beta^{\rm(tot)}(\eta,kR)
=\nu_{\beta}-\frac{\pi}{4}\epsilon
-\arg\left\{\erf\left({\cal Z}_{\beta,1}^\parallel,
{\cal Z}_{\beta,2}^\parallel\right)\right\}
\label{maslophasetot}
\end{equation}
that analytically connects the asymptotic values $\nu_\beta$ and
depends on the energy through $kR$. The final result
(\ref{maslophasetot}) for the total phase depends also on the
deformation parameter $\eta$.

Note that $\sigma^*$ is negative for $\eta\!<\!\eta_{\rm bif}$.  In the
derivation of Eqs.~(\ref{deltag1}) and (\ref{amp1}), we have changed
the integration variable from $\sigma$ to
$z\!=\!\sqrt{-i\epsilon|J_\beta^\parallel|/(2\hbar)}(\sigma-\sigma^*)$ in
order to transfer the $kR$ and $\eta$ dependence of the integrand to
the limits of the complex generalized error functions (\ref{errorf}).
Note also that our energy and deformation dependent phase
$\nu_\beta^{\rm(tot)}$ is essentially different from Ref.~\cite{bremen} and
much simpler in its analytical structure. Different from
Refs.~\cite{bremen,sie96}, we did not use any assumption concerning a
smoothness of the phase. Our solution is regular at the separatrix and
creeping points, at all bifurcations points and in the circular disk
limit. We easily get the correct circular disk limit~\cite{disk} and the
Berry-Tabor result~\cite{bt76} for larger deformations far from the
bifurcations.

Equations (\ref{deltag1}), (\ref{amp1}) and (\ref{maslophasetot})
represent one of our central results concerning the contributions of
the degenerate orbit families (${\cal K}=1$), that simultaneously
solves the symmetry-breaking problem for both hyperbolic and elliptic
orbits: near $\eta=1$ and other bifurcation points for all hyperbolic
orbits, and near the separatrix $\sigma_s$ and the `creeping' point
$\sigma_{\rm cr}$ for all elliptic orbits. The additional
contributions of the isolated orbits (${\cal K}=0$) will be derived in
the two following subsections.

Formally, our result (\ref{deltag1}) coincides with the first main term
of the Berry-Tabor trace formula, see Eq.~(24) of Ref.~\cite{bt76},
using the simplest way of the expansions near the stationary point
instead of a more general and more complicated mapping procedure. The
next two terms of their formula, being of higher order in
$\sqrt{\hbar}$, can be obtained by accounting for the linear term in
the expansion of the pre-exponential factor over $\sigma-\sigma^*$.
They were neglected in our approach because we are interested here only
in the main term of the SPM expansion, in order to get the simplest
possible solution of the bifurcation problem. With the higher-order
corrections, we should take into account that the ratio of the
contribution of the linear term to the zero-order term of the amplitude
is of the same order as the relative contribution of the next order
(cubic) term in the expansion of the phase.  For a consistent treatment
of the level density in the semiclassical asymptotic approximation $kR
\gg 1$, one would have to collect both corrections.

\subsection{Short diametric orbits $({\cal K}=0)$}

For the contribution of the isolated (${\cal K}=0$) diameters, only one
of the two stationary phase conditions (\ref{statcondt}) corresponding
to the $\Theta_v$ variable is an identity.  The other one for
$\Theta_u$ is a nontrivial equation for the discrete number of the
stationary points which differ by integer multiple of $\pi$. Indeed,
due to the integrability of motion in the elliptic billiard one has
\begin{equation}
\Theta_u = \omega_u t+\Theta_u^{(0)}, \qquad
\Theta_v = \omega_v t+\Theta_v^{(0)},
\label{angles}
\end{equation}
where $\bbox{\Theta}^{(0)}$ is the initial angle $\bbox{\Theta}$ at
$t=0$. Since the frequency $\omega_u$ in Eqs.~(\ref{angles}) is zero
for short diameters, for instance, there is no room for an identity in
the stationary phase condition for the variable $\Theta_u$ in
Eq.~(\ref{pstraceactang1}).  Hence, the Poisson-sum trace formula
cannot be applied to get the contribution from the short diameters
unlike in the derivations in Ref.~\cite{richens}. The stationary
points for the integration in Eq.~(\ref{pstraceactang1}) over angle
$\Theta_u$ for the short diameters are constants $\Theta_u^*=\pi M$ for
$M=0, \pm 1,\ldots$. Due to the periodicity of the angle variable with
the period $2\pi$ we really need to deal with the two stationary points
$\Theta_u^*=0$ and $\pi$ in the integration interval from $-\pi$ to
$\pi$ over the angle $\Theta_u$ in Eq.~(\ref{pstraceactang1}). We can
then reduce the initial integration interval for angle variable
$\Theta_u$ to the region from $-\pi/2$ to $\pi/2$ taking into account
the integration over other angles (related to the motion along the same
periodic orbit in the opposite direction) by the factor 2 (due to the
time reversal invariance of the Hamiltonian). Within this reduced
integration interval, only one stationary point $\Theta_u^*=0$ must be
taken into account in the calculation by the improved stationary phase
method.

For the other variable $\Theta_v$ for the short diameters we have
identity in the corresponding equation from Eq.~(\ref{statcondt}). The
integrand in (\ref{pstraceactang1}) is independent of the variable
$\Theta_v$ and the integral gives simply $2\pi$.  Thus, the integrand
for the contribution of the short diameters essentially depends only on
$\Theta_u$ and possesses the relevant stationary points. When we take
this integral by the SSPM we get immediately Gutzwiller's result for
the short diameters with his stability factor in the denominator.  This
stability factor is zero at the bifurcation points.  We shall below get
the short diameter term improved at the bifurcation points.  For this
purpose we shall first follow the same method in the integration over
$\Theta_u$ and $I_u$ as we did in the integration over $I_u$ for
elliptic and hyperbolic orbits with highest degeneracies. The
integration interval over $I_u$ for the contribution of the short
diameters is also finite from 0 to the maximal ``creeping'' value
$I_u^{\rm(cr)}$ (\ref{iucr}) which corresponds to the region of the
$\sigma$ variable $0 \leq\sigma \leq \sigma_{\rm cr}$.

Thus, for the short diameters, we use the stationary condition for the
angle variable $\Theta_u$ and expand the phase of exponent in
Eq.~(\ref{pstraceactang1}) about the short diameter,
\begin{equation}
S_\alpha = S_{sM}(\varepsilon)+\frac12 J_{sM}^\perp\Theta_u^2,
\label{actionexps}
\end{equation}
with $S_{sM}(\varepsilon)$ being the action along the short diameter,
$S_{sM}(\varepsilon)=4\,p(\varepsilon)\,aM$ and $\Theta_u^*=0$.
$J_{sM}^\perp$ is the Jacobian corresponding to the second variation of
the action $S_\alpha$ with respect to the angle variable $\Theta_u$,
\begin{eqnarray}
J_{sM}^\perp &=& \left(\frac{\partial^2 S_\alpha}{\partial\Theta_u'^2}
+ 2\frac{\partial^2 S_\alpha}{\partial\Theta_u'\partial\Theta_u''}
+ \frac{\partial^2 S_\alpha}{\partial\Theta_u''^2}\right)_{sM}
= \left(-\frac{\partial I_u'}{\partial \Theta_u'}
- 2\frac{\partial I_u'}{\partial \Theta_u''}
+ \frac{\partial I_u''}{\partial \Theta_u''}\right)_{sM},
\nonumber\\
\label{jacobperp}
\end{eqnarray}
according to Eq.~(\ref{phasecondti}).  The Jacobian $J_{sM}^\perp$ is
expressed in terms of the diametric curvature $K_s$ (\ref{curvshortd})
and Gutzwiller's stability factor $F_{sM}$,
\begin{eqnarray}
F_{sM} = -\left(\frac{
 -\frac{\partial I_u'}{\partial \Theta_u'}
 -2\frac{\partial I_u'}{\partial \Theta_u''}
 +\frac{\partial I_u''}{\partial \Theta_u''}}{
 \frac{\partial I_u'}{\partial \Theta_u''}}
\right)_{sM}
= 4\sin^2\left[M\arccos(2\eta^{-2}-1)\right],
\label{gutzstabfacts}
\end{eqnarray}
which is independent of the choice of the phase space variables
\begin{equation}
J_{sM}^\perp=F_{sM}J_{sM}^{(\Theta)}
= -\frac{F_{sM}}{4\pi MK_s},
\label{jacobperprelations}
\end{equation}
where
\begin{equation}
J_{sM}^{(\Theta)}=-\left(
\frac{\partial I_u'}{\partial\Theta_u''}\right)_{sM}
\label{jacobpars}
\end{equation}
and $K_s$ is the short diametric curvature given by
Eq.~(\ref{curvshortd}) ($\epsilon=-1$).  In the second equality of
Eq.~(\ref{jacobperprelations}) we used a simple relation between the
Jacobians $J_{sM}^{(\Theta)}$, $J_{\beta}^\parallel$ and $K_s$.  This
relation follows directly from their definitions and simple properties
of the Jacobians:
\begin{equation}
\frac{J_{sM}^{(\Theta)} J_{\beta}^\parallel}{
\left(\frac{\partial I_u}{\partial \sigma}\right)^2}=-1.
\label{jacobprop}
\end{equation}

After the exact integration over $\Theta_v$ in
Eq.~(\ref{pstraceactang1}) which gives $2\pi$ as explained above,
we substitute
the expansion (\ref{actionexps}) of the action $S_\alpha$ and take the
amplitude factor at the stationary point $\Theta_u^*=0$.  We take the
integral over $\Theta_u$ within the finite range from $-\pi/2$ to $\pi/2$
which can be reduced more to the integral from $0$ to $\pi/2$ with
the factor
$2$ due to the spatial symmetry in addition to the
time reversibility factor $2$ mentioned above.
Integrating over $I_u$ as in the previous
subsection, one finally gets
\begin{equation}
\delta g_{{\rm scl},0}^{(s)}
= \Re\sum_M {\cal A}_{sM}^{(0)}\exp[ikL_{sM}-i\nu_{sM}].
\label{deltagsd}
\end{equation}
Here, $L_{sM}$ is the length of the diameter orbit, $L_{sM}=4Ma$,
\begin{equation}
{\cal A}_{sM}^{(0)}
= \frac{2a}{\varepsilon_0\pi kR^2}\,
  \frac{1}{\sqrt{\left|F_{sM}\right|}}
\erf\left({\cal Z}_{sM,1}^\parallel,{\cal Z}_{sM,2}^\parallel\right)
\erf\left({\cal Z}_{sM,1}^\perp,{\cal Z}_{sM,2}^\perp\right),
\label{amps0}
\end{equation}
${\cal Z}_{sM,1}$ and ${\cal Z}_{sM,2}$ are defined by
\begin{equation}
{\cal Z}_{sM,1}^\parallel=0, \quad
{\cal Z}_{sM,2}^\parallel
= \sqrt{\frac{i\left|J_{sM}^\parallel\right|}{2\hbar}}\sigma_{\rm cr},
\label{argerrorpars}
\end{equation}
\begin{equation}
{\cal Z}_{sM,1}^\perp
= \sqrt{\frac{-i\left|J_{sM}^\perp\right|}{2\hbar}}\Theta_u'=0, \quad
{\cal Z}_{sM,2}^\perp
= \sqrt{\frac{-i\left|J_{sM}^\perp\right|}{2\hbar}}\Theta_u''
= \frac{\pi}{2}\sqrt{\frac{-i\left|J_{sM}^\perp\right|}{2\hbar}}.
\label{argerrorperp}
\end{equation}
For any finite deformation and sufficiently large $kR$,
Eq.~(\ref{amps0}) is much simplified
by using asymptotics for the first error function
and one gets
\begin{equation}
{\cal A}_{sM}^{(0)}
= \frac{2a}{\varepsilon_0\pi kR^2}
  \frac{1}{\sqrt{\left|F_{sM}\right|}}
\erf\left({\cal Z}_{sM,1}^\perp,{\cal Z}_{sM,2}^\perp\right).
\label{ampsd}
\end{equation}

The constant part $\nu_{sM}$ of the Maslov phases in
Eq.~(\ref{deltagsd}) is obtained in the same way as in the previous
subsection,
\begin{equation}
\nu_{sM}=3\pi M-\frac{\pi}{2}.
\label{phaseshortd}
\end{equation}

For deformations far from the bifurcation points, the level density
$\delta g_{{\rm scl},0}^{(s)}$ (\ref{deltagsd}) asymptotically reduces
to the standard Gutzwiller formula for isolated short
diameters,~\cite{gutzpr,gutz,book}
\begin{equation}
\delta g_{{\rm scl},0}^{(s)}(\varepsilon)\to
\frac{2a}{\varepsilon_0 \pi kR^2}\sum_M \frac{1}{\sqrt{F_{sM}}}
\sin(k L_{sM}-\nu_{sM}).
\label{deltagsdg}
\end{equation}
The total Maslov phase $\nu_{sM}^{\rm(tot)}$ for the diameter orbits is
\begin{eqnarray}
\nu_{sM}^{\rm(tot)}
&=& \nu_{sM}
- \arg\left\{\erf\left(
{\cal Z}_{1,sM}^{\parallel},{\cal Z}_{2,sM}^{\parallel}\right)\right\}
- \arg\left\{\erf\left(
{\cal Z}_{1,sM}^{\perp},{\cal Z}_{2,sM}^{\perp}\right)\right\}
\nonumber\\
&\approx& \nu_{sM}
-\arg\left\{\erf\left(
{\cal Z}_{1,sM}^{\perp},{\cal Z}_{2,sM}^{\perp}\right)\right\}
\label{maslovphasetotsd}
\end{eqnarray}
for large $kR$.

Near the bifurcation points where $F_{sM}\to 0$, one gets
from Eq.~(\ref{deltagsd}) the finite limit,
\begin{eqnarray}
\delta g_{{\rm scl},0}^{(s)} &\to& \frac{a}{\pi\varepsilon_0 kR^2}\,
\Re\sum_M\frac{1}{\sqrt{2Mi\hbar\left|K_s\right|}}
\erf\left({\cal Z}_{sM,1}^{\parallel},{\cal Z}_{sM,2}^{\parallel}\right)
e^{i(kL_{sM}-\nu_{sM})} \nonumber\\
&\approx& \frac{\eta^{1/4}}{\varepsilon_0 \sqrt{2\pi kR}}
\Re\sum_M\frac{1}{\sqrt{M}}e^{i(kL_{sM}-\nu_{sM}-\pi/4)}.
\label{deltagsbif}
\end{eqnarray}
Note that the two last terms in Eq.~(24) of Ref.~\cite{bt76} are
smaller than the above contribution (\ref{deltagsbif}) at the
bifurcation deformations $\eta_{\rm bif}$ (\ref{etamin}) by the factor
$\sqrt{kR}$.  Therefore, these two terms are the next order
semiclassical corrections and can be neglected compared to the term
(\ref{deltagsbif}) obtained above. Moreover, the ISPM solution
(\ref{deltagsd}) is not related to the ``diametric'' part of the
Poisson-sum trace formula (\ref{poissonsums}) with $n_u=1,n_v=2$ as
follows from the derivations in Ref.~\cite{richens}
($\alpha_1=2,\alpha_2=\lambda=2$ in the notations of
Ref.~\cite{richens} applied for the short diameters in the elliptic
billiard, $\alpha_1=2n_u$) (see a more detailed discussion below).
Thus, our derivation is essentially different from that suggested
earlier in Ref.~\cite{richens} (where the last two terms in Eq.~(24)
of Ref.~\cite{bt76} are retained without considering the contribution
(\ref{deltagsbif})).

Taking the limit of Eq.~(\ref{deltagsbif}) for $\eta \rightarrow 1$ we
obtain the same contribution of the diameters in the circular
disk~\cite{disk} as found from the ``diametric'' part
of the Poisson-sum trace formula,
\begin{equation}
\delta g_{{\rm scl},1}^{(d)}(\varepsilon)
= \frac{1}{\varepsilon_0 \sqrt{2\pi kR}}\sum_M\frac{1}{\sqrt{M}}
\sin(k L_{sM}-\nu_{sM}+\pi/4).
\label{diskd}
\end{equation}
The value at this limit is larger by the factor $\sqrt{kR}$ than the
standard Gutzwiller result for isolated orbits like at any other
bifurcation points.

\subsection{Long diameters and the separatrix}

As shown in \S2, the curvature $K$ goes to infinity being positive
from the right side and negative from the left side near the separatrix
($\sigma=1$) with the same modulus, see Eqs.~(\ref{curvate}),
(\ref{curvath}) and Fig.~\ref{fig2}(b).  The derivation for short
diameters of the previous section with the expansion of the action
exponent phase to second order terms cannot be applied in this case.
However, we note that the behaviour of the curvature near the
separatrix in the action $I_u$ (or $\sigma$) variable is similar to
that for the eigenvalues of the matrix of the second derivatives of
the action in the usual coordinate space near the turning points.  One
can thus apply the Maslov and Fedoryuk idea for the calculation of the
Maslov indices, see Refs.~\cite{fed:jvmp,masl,fed:spm,masl:fed}.
Following this idea we expand first the phase of exponent in
Eq.~(\ref{pstraceactang}) with respect to the action $I_u$ taking into
account the next third order terms, see Eq.~(\ref{expan3}) in
Appendix~B.  We then use the linear transformation (\ref{alphabeta})
to the new variable $z$ to get the standard exponent in the integral
representation of the Airy functions.  Within this method we take a
small first derivative (small parameter $c_1$) and the large second
derivative (curvature) in the cubic polynomial expansions
(\ref{expan3}) taking $\sigma $ on the small distance from the
separatrix $\sigma=1$.  After some algebraic transformations we obtain
Eq.~(\ref{pstracelang}) in Appendix~B for the limit $\sigma
\rightarrow 1$. Note that a similar idea which we used here
considering $\sigma$ {\em near} the singular separatrix point
$\sigma=1$ and only finally, after the calculation of the integrals,
put $\sigma \rightarrow 1$ was applied in the derivations of the {\em
separate} contributions of the hyperbolic orbit family and short
diameters to the periodic orbit sum, as mentioned above.

For the angle integral in Eq.~(\ref{pstracelang}) we use the same
Maslov-Fedoryuk method~\cite{fed:jvmp,masl,fed:spm,masl:fed} applied
for the caustic case. As result, one obtains (see Appendix~B)
\begin{eqnarray}
\delta g_{{\rm scl},0}^{(l)}(\varepsilon)
&=& \frac{b}{\varepsilon_0 \pi kR^2}\Re\sum_M
e^{i[kL_{lM}+\frac23(w_\parallel^{3/2}+w_\perp^{3/2})-\nu_{lM}]}
\nonumber \\
&& \times \sqrt{\frac{\sqrt{w_\parallel w_\perp}}{
\left|c_2^\parallel c_2^\perp\right|}}
\bigl[\Ai(-w_\parallel)+i\Gi(-w_\parallel)\bigr]
\nonumber \\
&& \times \bigl[\Ai\left(-w_\perp,
{\cal Z}_{lM,1}^\perp,{\cal Z}_{lM,2}^\perp\right)
+i\Gi\left(-w_\perp,
{\cal Z}_{lM,1}^\perp,{\cal Z}_{lM,2}^\perp\right)\bigr].
\label{deltagld}
\end{eqnarray}
Here, the complete and incomplete Airy (or Gairy) functions with one and
three arguments (Eq.~(\ref{airynoncomp})) are used in line of the
definitions in Refs.~\cite{abramov,frahn}, see also Appendix~B for the
definitions of all other quantities.

For large $kR\sqrt{\eta^2-1}$, near the separatrix $\sigma\to 1$ the
parameter $w_\perp$ is negligible in Eq.~(\ref{zmimalperp}) for the limits
${\cal Z}_{1,lM}^\perp$ and ${\cal Z}_{2,lM}^\perp$ and the
integration range can be extended from $0$ to $\infty$.  The
incomplete Airy integrals in Eq.~(\ref{deltagld}) approach the
complete ones and the asymptotics of all Airy functions like $\Ai(-w)$
and $\Gi(-w)$ is now applied~\cite{abramov}.  Finally, we get
asymptotically the standard Gutzwiller result for the isolated
diameters,~\cite{gutzpr,gutz,book}
\begin{eqnarray}
\delta g_{{\rm scl},0}^{(l)}(\varepsilon)
&=& -\frac{2b}{\varepsilon_0 kR^2}\Re\sum_M
e^{i[kL_{lM}+\frac23(w_\parallel^{3/2}+w_\perp^{3/2})-\nu_{lM}]}
\sqrt{\frac{\sqrt{w_\parallel w_\perp}}{\left|F_{lM}\right|}}
\nonumber \\
&& \times
\bigl[\Ai(-w_\parallel)+i\Gi(-w_\parallel)\bigr]
\bigl[\Ai(-w_\perp)+i\Gi(-w_\perp)\bigr] \nonumber \\
&\to& \frac{2b}{\varepsilon_0 \pi kR^2}
\sum_M \frac{1}{\sqrt{\left|F_{lM}\right|}}
\sin(kL_{lM}-\nu_{lM})
\label{deltagldg}
\end{eqnarray}
where $F_{lM}$ is the Gutzwiller stability factor for long diameters,
\begin{equation}
F_{lM} = -4\sinh^2\left[M\arccosh(2\eta^2-1)\right],
\label{gutzstabfactl}
\end{equation}
\begin{equation}
\nu_{lM}=3\pi M-\frac{\pi}{2}.
\label{phaseld}
\end{equation}
In the second equation we used the asymptotics of the $\Ai(-w)$ and
$\Gi(-w)$ functions~\cite{abramov}.  We found also the constant
part $\nu_{lM}$ of the phase by using the Maslov-Fedoryuk theory.  The
deformation and energy dependent Maslov phases are determined by the
additional phases in the exponent and the argument of the product of
the square brackets in (\ref{deltagld}) through complex combinations
of the Airy and Gairy functions and their arguments.

In the circular shape limit both the upper and the lower limits of the
incomplete Airy functions in Eq.~(\ref{deltagld}) tend to zero and the
angle integral has the finite limit $\pi/2$ because $c_2^\parallel$,
$c_3^\perp$ and $w_\perp$ vanish, see Appendix~B. With this, the other
factors near the separatrix $\sigma\to 1$ ensure that the amplitudes
for long diameters diminish because $w_\parallel$ (\ref{wpar}) vanishes
at the separatrix, see also Ref.~\cite{abramov}.  Namely, the long
diameter contribution becomes zero in the circular shape limit.

Thus, for deformations far from the bifurcations, the results
(\ref{deltagsd}) and (\ref{deltagld}) of the ISPM reduce to the
standard Gutzwiller formula.  In the circular disk limit the improved
short diameter density (\ref{deltagsd}) approaches continuously the
diametric contribution to the circular disk density, while the long
diameter (separatrix) contribution diminishes. Note that our ISPM
solution (\ref{deltagld}) for the unstable long diameters is not
related to the Poisson-sum trace formula (\ref{poissonsum}),
in particular, with its ``diametric'' part because of the
existence of the {\em isolated} stationary points for the {\em angle}
variable $\Theta_u$ like for the short diameters. Moreover, the
uniform approximation Eq.~(24) of Ref.~\cite{bt76} is singular at the
separatrix because of the divergence of the curvature $K_l$ for
$\sigma \rightarrow 1$, as noted in Ref.~\cite{bremen}.  However,
instead of the suggestion of Ref.~\cite{bremen} to use the
continuation of the WKB approach to the complex plane we applied
simpler Maslov-Fedoryuk method~\cite{fed:jvmp,masl,fed:spm,masl:fed}
and got the analytical dependence of the Maslov phase on the
deformation and energy through the exponent phase and complex
arguments of the Airy functions as well as their complex
summations.

\subsection{Closed orbits and the circular disk limit}
\label{closedorbitsdisk}

\begin{figure}
\chkfig{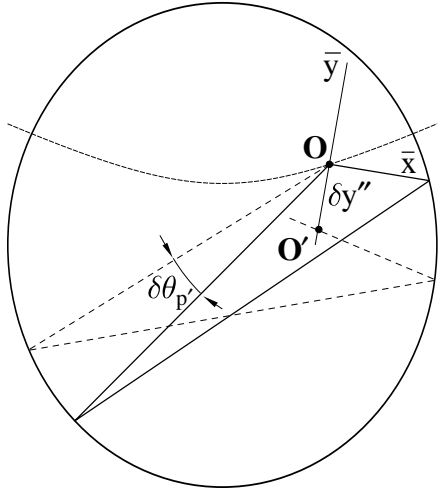}
\iffigs
\epsfxsize=.4\textwidth\centerline{\epsffile{fig03.ps}}
\else
\vspace{73mm}
\fi
\caption{\label{fig3}
Illustration of the caustic method for evaluating the
stability factor $ J_\alpha$ in Eq.~(\ref{jacobianalpha}) for the
closed two-reflection orbit ``co2''. The deflection
angle $\delta\theta'_p$ at the initial point $O(\boldr')$, variation
$\delta\bar{y}''$ of the final point $O'(\boldr'')$ with respect to $O$
and the coordinate system $(\bar{x},\bar{y})$ are shown:
Thick solid lines and dashed lines show the hyperbolic orbit ``co2'' and
the perturbed orbit, respectively;
the thin solid curve indicates the orbit-length invariant hyperbola
confocal to the boundary.}
\end{figure}

To get a more exact solution for the diameter contribution to the level
density and check the precision of the ISPM, we come back to the
initial trace formula Eq.~(\ref{trace1}) before application of the
ISPM for the calculation of this trace.\footnote{Eq.~(\ref{trace}) can
be obtained also from the phase space trace formula
Eq.~(\ref{pstrace}) taking the integral over two components of the
momentum $\boldp'$ along the energy surface by the stationary phase
method.}  For this purpose we shall take exactly the trace integral
(\ref{trace1}) in suitable variables.  This is the trace formula in
terms of the sum over all closed (periodic and non-periodic) orbits
$\alpha$,
\begin{equation}
\delta g_{\rm scl}(\varepsilon)
= 2\left(2\pi\hbar\right)^{-3/2}\frac{m}{\sqrt{p}}
\sum_\alpha\int\frac{dx\,dy}{\sqrt{J_{\alpha}(x,y)}}
\sin(kL_\alpha-\nu_\alpha),
\label{trace2}
\end{equation}
where $J_{\alpha}(x,y)$ is the stability factor defined through the
Jacobian ${\cal J}_\alpha(\boldp't_\alpha,\boldr''\varepsilon)$ by
\begin{equation}
{\cal J}_\alpha(\boldp't_\alpha,\boldr''\varepsilon)
= \frac{m^2}{p}\left(\frac{\partial\theta_p'}{\partial{\bar y}''}
\right)_\alpha
= \frac{m^2}{p}\frac{1}{J_\alpha(x,y)}.
\label{jacobianalpha}
\end{equation}
Here the deflection $\delta{\bar y}''$ of the final path point in the
perpendicular direction of the local Cartesian system $({\bar x},{\bar
y})$ comes from the angle variation $\delta\theta_p'$ of the initial
momentum,~\cite{mfimbrk,disk} see Fig.~\ref{fig3}.

\begin{figure}
\chkfig{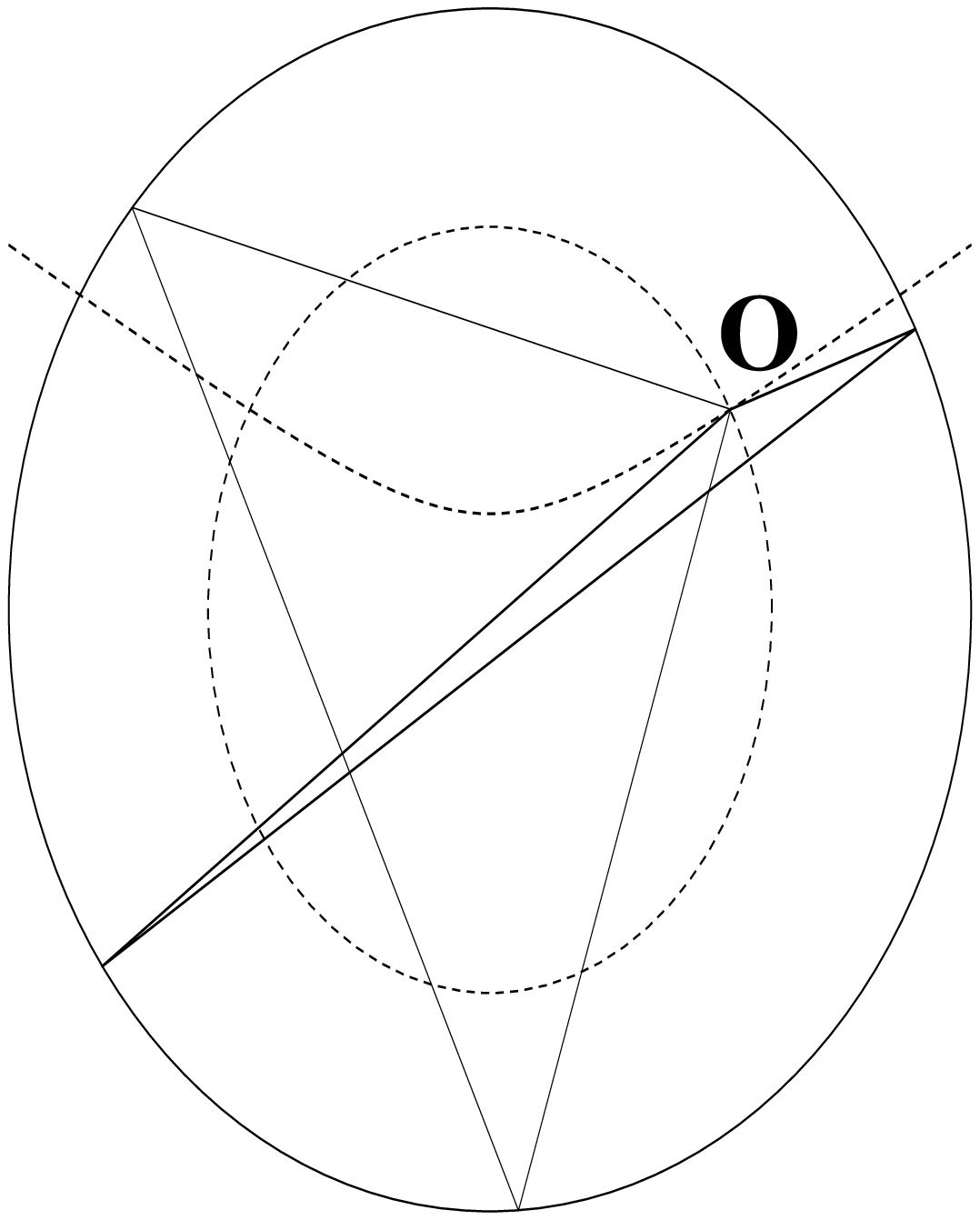}
\iffigs
\begin{center}
\begin{minipage}{.35\textwidth}
\epsfxsize=.6\textwidth\centerline{\epsffile{fig04a.ps}}
\end{minipage}
\begin{minipage}{.35\textwidth}
\epsfxsize=.6\textwidth\centerline{\epsffile{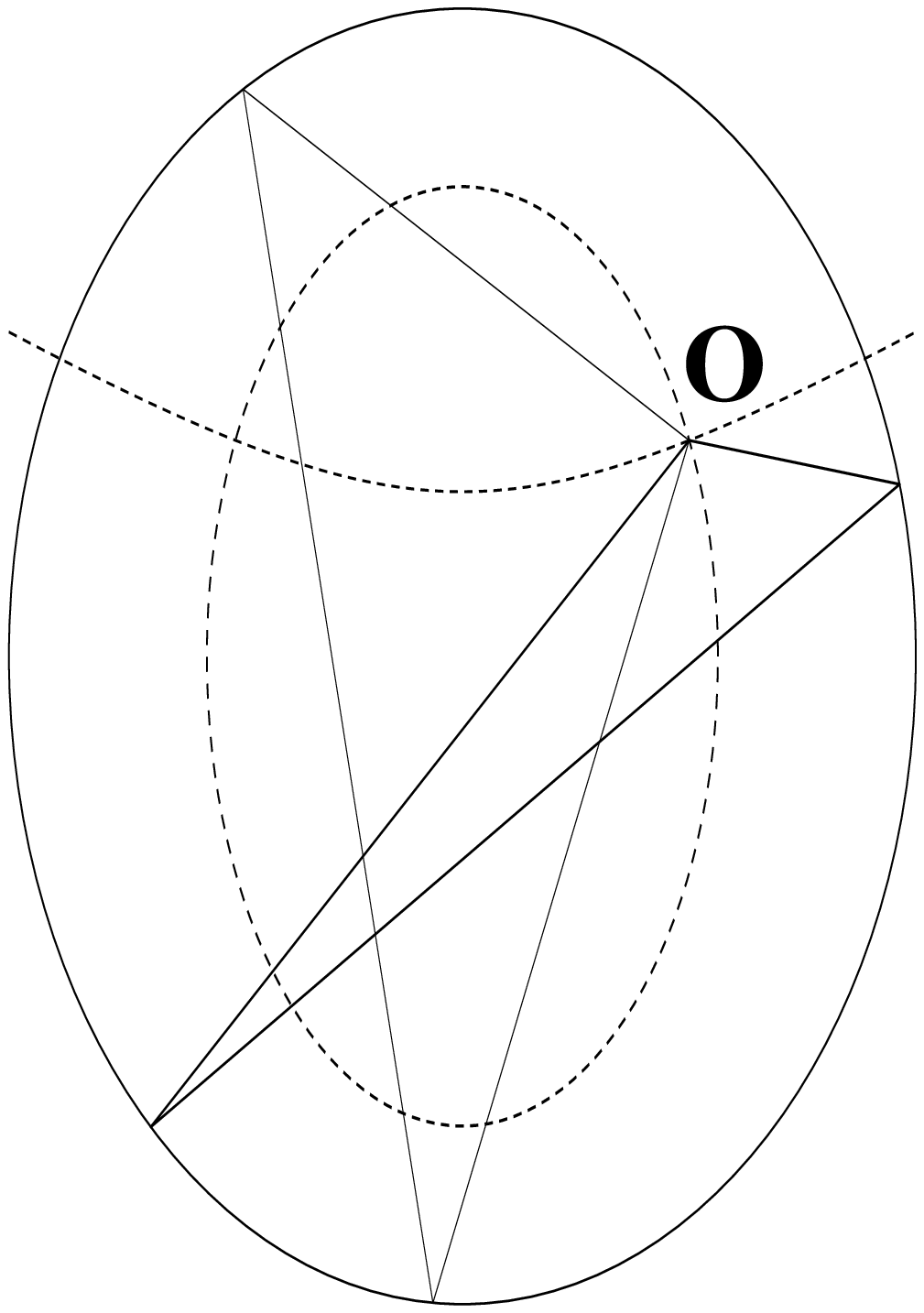}}
\end{minipage}
\end{center}
\else
\vspace{48mm}
\fi
\caption{\label{fig4}
Closed non-periodic two-reflection orbits with the
elliptic and hyperbolic caustics at the initial point $O(x,y)$ are
shown by thin and thick solid lines, respectively,
for the deformation $\eta=1.05$ (left-hand side) and $1.2$ (right-hand
side): $O$ is the common vertex of both triangular orbits; dashed curves
indicate the orbit-length invariant ellipse and hyperbola crossing the
initial point. The hyperbolic orbit is close to the diameter of the
circular shape for small deformations.}
\end{figure}

We shall then simplify the trace formula (\ref{trace2}) taking the
contribution of the main shortest closed orbits $\alpha$ with the two
reflection points denoted below by the index ``co2'' as an example. For
arbitrary point $(x,y)$ inside the elliptic billiard one can find such
orbits ``co2'' which are the triangles with the two vertices at the
elliptic boundary and one vertex at the point $(x,y)$, see
Fig.~\ref{fig4}.
There are two kind of such orbits.  For any point
$(x,y)$ we can plot the hyperbola and ellipse confocal to the boundary,
which are the orbit-length invariant curves.  Indeed, moving the
initial point $(x,y)$ along such hyperbola (or ellipse) we have the
one-parametric family of the triangle-like orbits with the same action
(${\cal K}=1$).  We shall call them for short the hyperbolic and
elliptic ``co2'' orbits, respectively.

For the calculation of the trace integral (\ref{trace2}) it is
convenient to use the elliptic coordinates $(u,v)$, (\ref{ellcoord}).
After this coordinate transformation, we can take the sine function of
the action off the $v$ or $u$ integration for the hyperbolic or
elliptic ``co2'' orbits, respectively, because of independence of the
action on the corresponding elliptic coordinate.  Finally, one obtains
from Eq.~(\ref{trace2})
\begin{eqnarray}
\delta g_{{\rm scl},1}^{\rm(hco2)}(\varepsilon)
= 2(2\pi\hbar)^{-3/2} \frac{m\zeta^2}{\sqrt{p}}
\int\frac{du\sin(kL_{\rm hco2}(u)-\nu_{\rm hco2})\,dv(\sinh^2v+\cos^2u)}{
\sqrt{J_{\rm hco2}(x(u,v),y(u,v))}}
\nonumber \\
\label{deltaghco2}
\end{eqnarray}
for the contribution from the hyperbolic ``co2'' orbits (hco2),
and a similar
equation for the elliptic ``co2'' orbits.
An explicit expression for the stability factor
$J_{\rm co2}(x,y)$ evaluated
by the caustic method~\cite{mfimbrk} is presented in Appendix C.

\begin{figure}
\chkfig{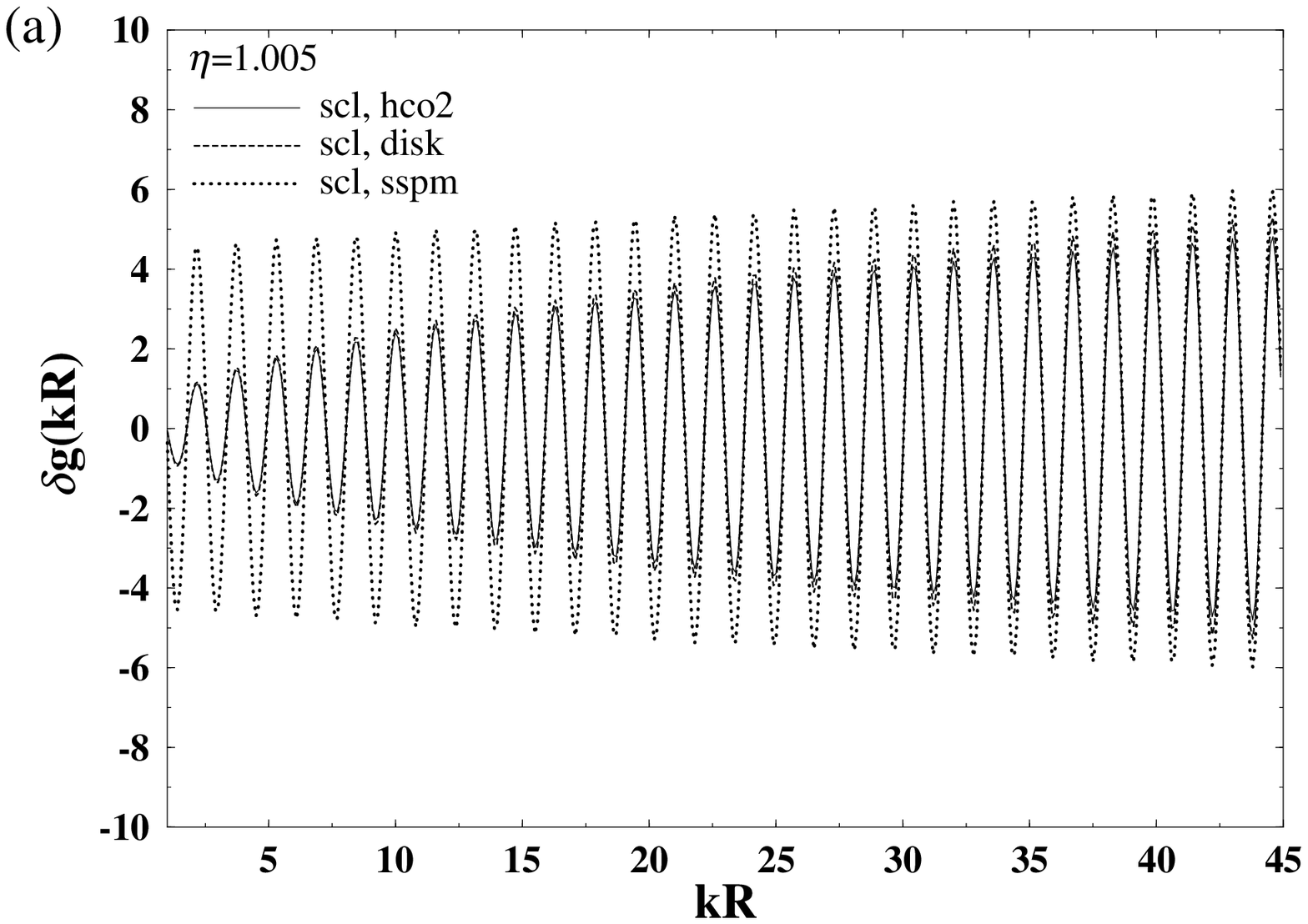}
\iffigs
\noindent
\begin{minipage}{.48\textwidth}
\epsfxsize=\textwidth\centerline{\epsffile{fig05a.ps}}
\end{minipage}\hfill
\begin{minipage}{.48\textwidth}
\epsfxsize=\textwidth\centerline{\epsffile{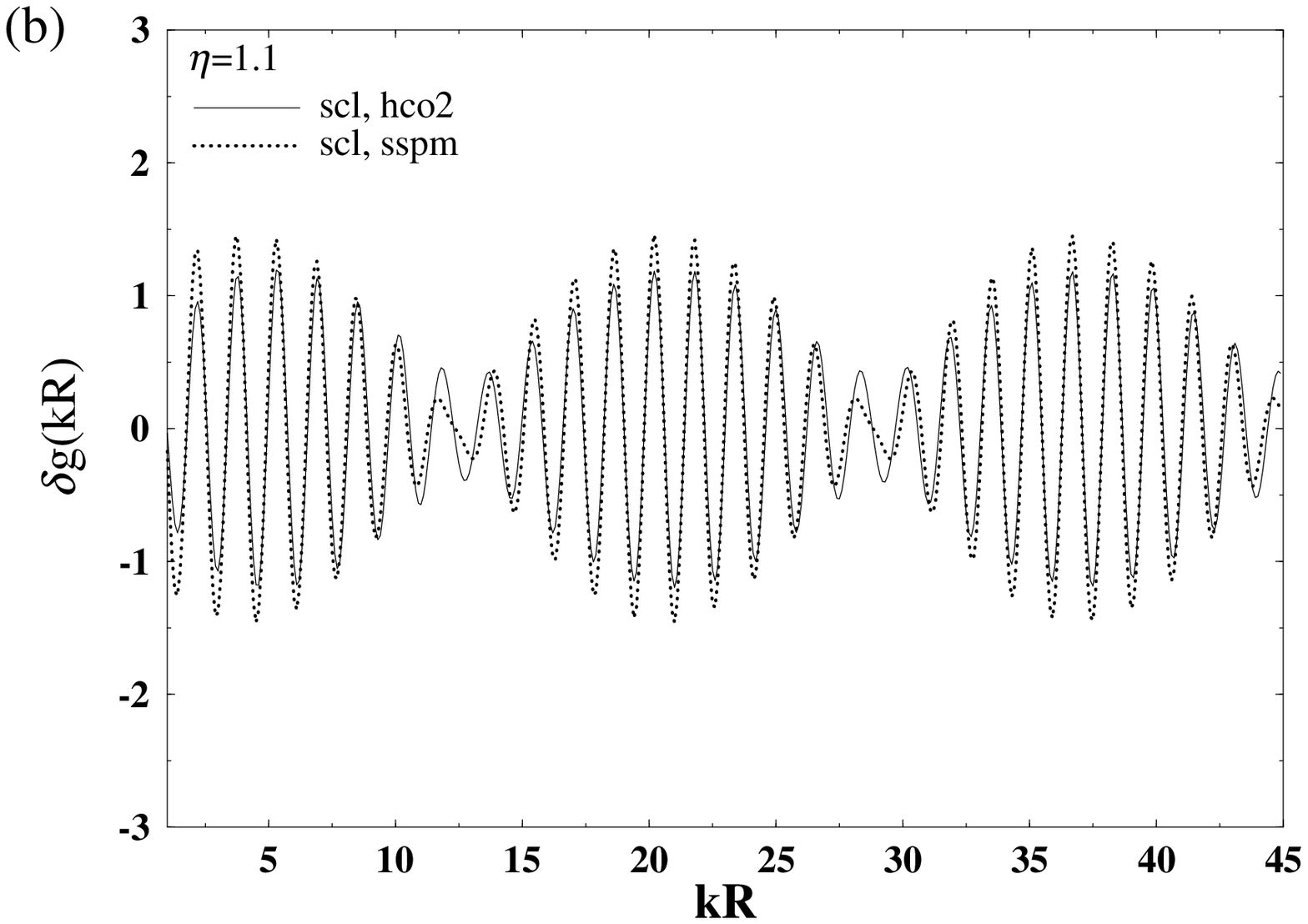}}
\end{minipage}
\else
\vspace{56mm}
\fi
\caption{\label{fig5}
(a) Convergence to the circular shape limit: The contribution of the closed
two-reflection orbits of the hyperbolic type ``hco2'' (see
Fig.~\ref{fig4}) to the level density $\delta g(kR)$
is shown by a solid line for the deformation $\eta=1.005$, while
Gutzwiller's trace formula (SSPM) for isolated diameters
and circular disk trace formula
are indicated by dotted and dashed lines,
respectively. The dashed line overlaps with the solid line, so that it
cannot be distinguished from the latter.
(b) Convergence to the Gutzwiller trace formula for
$\eta=1.1$. Notations are the same as in (a).}
\end{figure}

Note that the hyperbolic ``co2'' orbits with the initial point $(x,y)$
reduce to the disk diameters crossing the same point in the circular
disk limit, see Fig.~\ref{fig4}.  The stability factor $J_{hco2}(x,y)$,
(\ref{jacobco2}), turns into the analytical circular disk expression of
Ref.~\cite{disk}.  The circular disk limit of the level density
(\ref{deltaghco2}) coincides with the diameter contribution $\delta
g_{{\rm scl},1}^{(d)}(\varepsilon)$, (\ref{diskd}), as shown in
Fig.~\ref{fig5}(a).  The opposite limit of (\ref{deltaghco2}) far from
the bifurcations is the Gutzwiller SPM for the short and long isolated
diameters, see Fig.~\ref{fig5}(b).  The contribution of the elliptic
``co2'' is negligibly small everywhere, and vanishes at the circular
disk shape as next order $\hbar$ corrections.

\section{Level Density, Shell Energy and Averaging}
\label{avdensesc}

\subsection{Total level density}
\label{totlevdens}

The total semiclassical POT density can be written as the sum over all
periodic orbit families considered in the previous section,
\begin{equation}
\delta g_{\rm scl}(\varepsilon)
= \delta g_{{\rm scl},1}(\varepsilon)
 +\delta g_{{\rm scl},0}^{(s)}(\varepsilon)
 +\delta g_{{\rm scl},0}^{(l)}(\varepsilon)
= \sum_\beta \delta g_{\rm scl}^{(\beta)}(\varepsilon),
\label{tracetotal}
\end{equation}
where the first term is the contribution (\ref{deltag1}) from the
elliptic and hyperbolic orbits.  The second and third terms are the
contributions from the short (\ref{deltagsd}) and the long
(\ref{deltagld}) diameters, respectively.  Near the circular limit, the
last two terms for one period ($M=1$) can be replaced by the
contribution of the hyperbolic ``co2'' orbits (\ref{deltaghco2}) to get
a more precise semiclassical result.

\subsection{Semiclassical shell energy}
\label{enshellcorr}

The shell-correction energy $\delta E$ can be expressed in terms of the
oscillating part $\delta g_{\rm scl}^{(\beta)}(\varepsilon )$ of the
semiclassical level density as~\cite{strusem,book,mfimbrk}
\begin{equation}
\delta E = 2\sum_\beta \left(\frac{\hbar}{t_\beta}\right)^2
\delta g_{\rm scl}^{(\beta)}(\varepsilon_F), \qquad
N=2\int_0^{\varepsilon_F} d\varepsilon g(\varepsilon).
\label{descl1}
\end{equation}
Here, $t_\beta$ is the time of the motion along the periodic orbit
$\beta$ (including its repetitions),
\begin{equation} t_\beta = M_\beta
T_\beta = \frac{2\pi M_\beta}{\Omega_\beta},
\label{tbeta}
\end{equation}
where $T_\beta$ is the period of primitive orbit with the Fermi energy
$\varepsilon_F$, $M_\beta$ the repetition number, $\Omega_\beta$ the
frequency, and $N$ the particle number. Note that we have taken into
account the spin degeneracy factor 2 in (\ref{descl1}).

The semiclassical representation of shell-correction energy
(\ref{descl1}) differs from that of $\delta g$ only by a factor
$(\hbar/t_\beta)^2=(\hbar^2k_F/mL_\beta)^2$, which suppresses
contributions from longer orbits. Thus short periodic orbits play
dominant roles in determining the shell-correction energy.

\subsection{Average level density}
\label{averdensity}

For the purpose of the presentation of the level density improved at
the bifurcation points we need to consider a level density averaged
slightly, thus avoiding the convergence problems that usually arise
when one is interested in a full semiclassical quantization.

The averaging is done by folding the level density with
a Gaussian of width $\Gamma$:
\begin{equation}
g_\Gamma(\varepsilon)
= \frac{1}{\sqrt{\pi}\Gamma}\int_{-\infty}^\infty
d\varepsilon'\,g(\varepsilon')\,
e^{-\left(\frac{\varepsilon-\varepsilon'}{\Gamma}\right)^2}.
\label{folding}
\end{equation}
The choice of the Gaussian form of the averaging function is immaterial
and guided only by mathematical simplicity. For cavities it is
convenient to use also the level density defined as a function of $kR$
averaged with a Gaussian of width $\gamma$:
\begin{equation}
g_\gamma(kR)
= \frac{1}{\sqrt{\pi}\gamma}\int_{-\infty}^{\infty}
d(k'R)\,g(k'R)\,e^{-\left(\frac{(k-k')R}{\gamma}\right)^2},
\label{folding2}
\end{equation}
where
\begin{equation}
g(kR)=\sum_i\delta((k-k_i)R)
=2kR\varepsilon_0\sum_i\delta(\varepsilon-\varepsilon_i)
=2kR\varepsilon_0g(\varepsilon),
\label{gkr}
\end{equation}
$\varepsilon_0=\hbar^2/2mR^2$
and the dimensionless parameter $\gamma$ is related to $\Gamma$ by
\begin{equation}
\Gamma=2\gamma\sqrt{\varepsilon\varepsilon_0}.
\label{gammas}
\end{equation}

Applying the averaging procedure defined above to the semiclassical
level density (\ref{tracetotal}), one gets~\cite{bablo,disk,mfimbrk}
\begin{equation}
\delta g_{\Gamma,scl}(\varepsilon)
= \sum_\beta \delta g_{\rm scl}^{(\beta)}(\varepsilon)\,
e^{-\left(\frac{\Gamma t_{\beta}}{2\hbar}\right)^2}
= \sum_\beta \delta g_{\rm scl}^{(\beta)}(\varepsilon)\,
e^{-\left(\frac{\gamma L_\beta}{2R}\right)^2}.
\label{dgsclgamma}
\end{equation}
The latter equation is written specifically for billiard problems in
terms of the orbit length $L_\beta$ (in units of a typical length
scale $R$) and $\gamma$.  The averaging yields an exponential decrease
of the amplitudes with increasing $L_{\beta}$ and/or $\gamma $.  As
shown in Ref.~\cite{mfimbrk}, for $\gamma$ of the order of unity, all
longer paths are strongly damped and only the shortest periodic orbits
contribute to the oscillating part of the level density, yielding its
gross-shell structure.  For a study of the bifurcation phenomenon,
however, we need smaller values of $\gamma$.

Finally, we should note that the higher the degeneracy of an orbit,
the larger the volume occupied by the orbit family in the phase space
and also, the shorter its length, the more important its
contribution to the average level density.


\section{Quantum Elliptic Billiard}

\subsection{Numerical method for the spectrum calculation}
\label{qmfourierspect}

Single-particle energies $\varepsilon_i$ of a particle of mass $m$
moving freely inside the elliptic boundary $v\leq v_b$ can be obtained
by a number of numerical methods.  Following the procedure employed in
previous works~\cite{ak95,ak97} by some of the present authors, one can
expand the deformed single-particle wave functions $\Psi(r,\theta)$
into a circular basis with well-defined orbital angular momentum $l$:
\begin{eqnarray}
\Psi^{(++)}_i(r,\theta)
&=& \sum_{l=0}^{(e)} A_l J_l(k_ir)\cos(l\theta), \quad
\Psi^{(-+)}_i(r,\theta)
= \sum_{l=1}^{(o)} B_l J_l(k_ir)\sin(l\theta),
\nonumber\\
\Psi^{(+-)}_i(r,\theta)
&=& \sum_{l=1}^{(o)} A_l J_l(k_ir)\cos(l\theta), \quad
\Psi^{(--)}_i(r,\theta)
= \sum_{l=2}^{(e)} B_l J_l(k_ir)\sin(l\theta),
\label{wave}
\end{eqnarray}
where $J_l(x)$ are the cylindrical Bessel functions of the first kind,
$k_i=\sqrt{2m\varepsilon_i}/\hbar$, the superscripts $(++)$ etc. stand
for parities with respect to reflections about the $x$ and $y$ axes,
and the superscripts $(e)$ and $(o)$ indicate the sums with respect to
even and odd $l$, respectively.  The expansion coefficients $A_l$ and
$B_l$ can be determined by applying Dirichlet boundary conditions.

In the present analysis, we also employed, in addition to the above
circular-wave decomposition method, the numerical procedure based on a
rather standard approach, the separation of the Schr\"odinger equation
in the elliptic coordinate system~\cite{bremen,morse53,misu97}.  In
terms of the elliptic coordinates (\ref{ellcoord}), the Schr\"odinger
equation can be written as
\begin{eqnarray}
\Bigl[\sqrt{\xi^2-1}\frac{\partial}{\partial\xi}
\left\{\sqrt{\xi^2-1}\frac{\partial}{\partial\xi}\right\}
&+& \sqrt{1-\phi^2}\frac{\partial}{\partial\phi}
\left\{\sqrt{1-\phi^2}\frac{\partial}{\partial\phi}\right\}\Bigr]
\psi(\xi,\phi) \nonumber\\
&+& \frac{2m\varepsilon_i \zeta^2(\xi^2-\phi^2)}{\hbar^2}
\psi(\xi,\phi) = 0,
\label{schrod}
\end{eqnarray}
where $\xi=\cosh v$ and $\phi=\cos u$.  Following Ref.~\cite{morse53},
this equation can be separated into two ordinary differential equations
by assuming $\psi(\xi,\phi)=R(\xi)S(\phi)$. The functions $R$ and $S$
are solutions of the ordinary differential equations
\begin{eqnarray}
(\xi^2-1)\frac{d^2R_l(c,\xi)}{d\xi^2}+\xi\frac{dR_l(c,\xi)}{d\xi}
-\left[\lambda_l-c^2\xi^2\right]R_l(c,\xi) &=& 0, \nonumber\\
(1-\phi^2)\frac{d^2S_l(c,\phi)}{d\phi^2}-\phi\frac{dS_l(c,\phi)}{d\phi}
+\left[\lambda_l-c^2\phi^2\right]S_l(c,\phi) &=& 0, \label{schrodsep}
\end{eqnarray}
where $\lambda_l$ is the separation constant and
$c=\zeta\sqrt{2m\varepsilon_i}/\hbar$ for $\xi\le\xi_b=\cosh v_b$. The
internal radial functions $R_l(c,\xi)$ are expanded in terms of Bessel
functions of the first kind. The expansion coefficients and the
separation constant $\lambda_l$ can be determined from the three-term
recurrence relations found in various
references~\cite{abramov,morse53,misu97,merch94}.

By imposing usual boundary conditions on the radial wave functions,
i.e., $R_l(c,\xi_b)=0$, one finds the eigenenergies $\varepsilon_i$.
All eigenvalues up to $kR\approx 40$ with the coordinate-transformation
method can be calculated numerically in matter of minutes without
overlooking solutions near level crossings, and hence the procedure is
certainly effective for the present model. The results obtained from
both numerical procedures were carefully compared and found to achieve
a nice convergence.

\begin{figure}
\chkfig{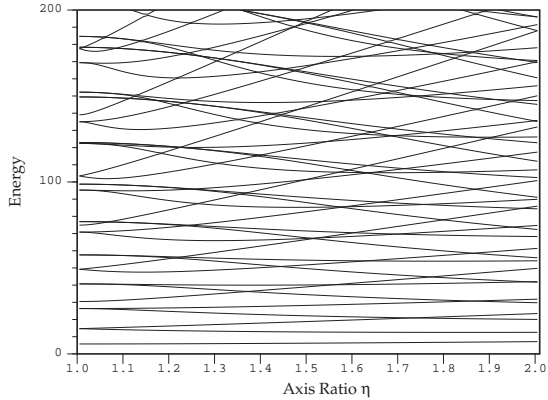}
\iffigs
\epsfxsize=.5\textwidth
\centerline{\epsffile{fig06.ps}}
\else
\vspace{54mm}
\fi
\caption{\label{fig6}
Single-particle spectra (in unit of $\varepsilon_0$)
for the elliptic billiard plotted as functions
of the deformation parameter $\eta$.}
\end{figure}

In Fig.~\ref{fig6} the deformation dependence of the single-particle
energies for the elliptic billiard is presented.  At the circular
limit, the familiar shell gaps are clearly observed, while different
shell gaps start to develop at higher deformations.  Below we identify
the semiclassical origin of these shell structures at higher
deformations.

\subsection{Strutinsky's smoothed level densities and shell energies}
\label{qmlevdens}

\begin{figure}[p]
\chkfig{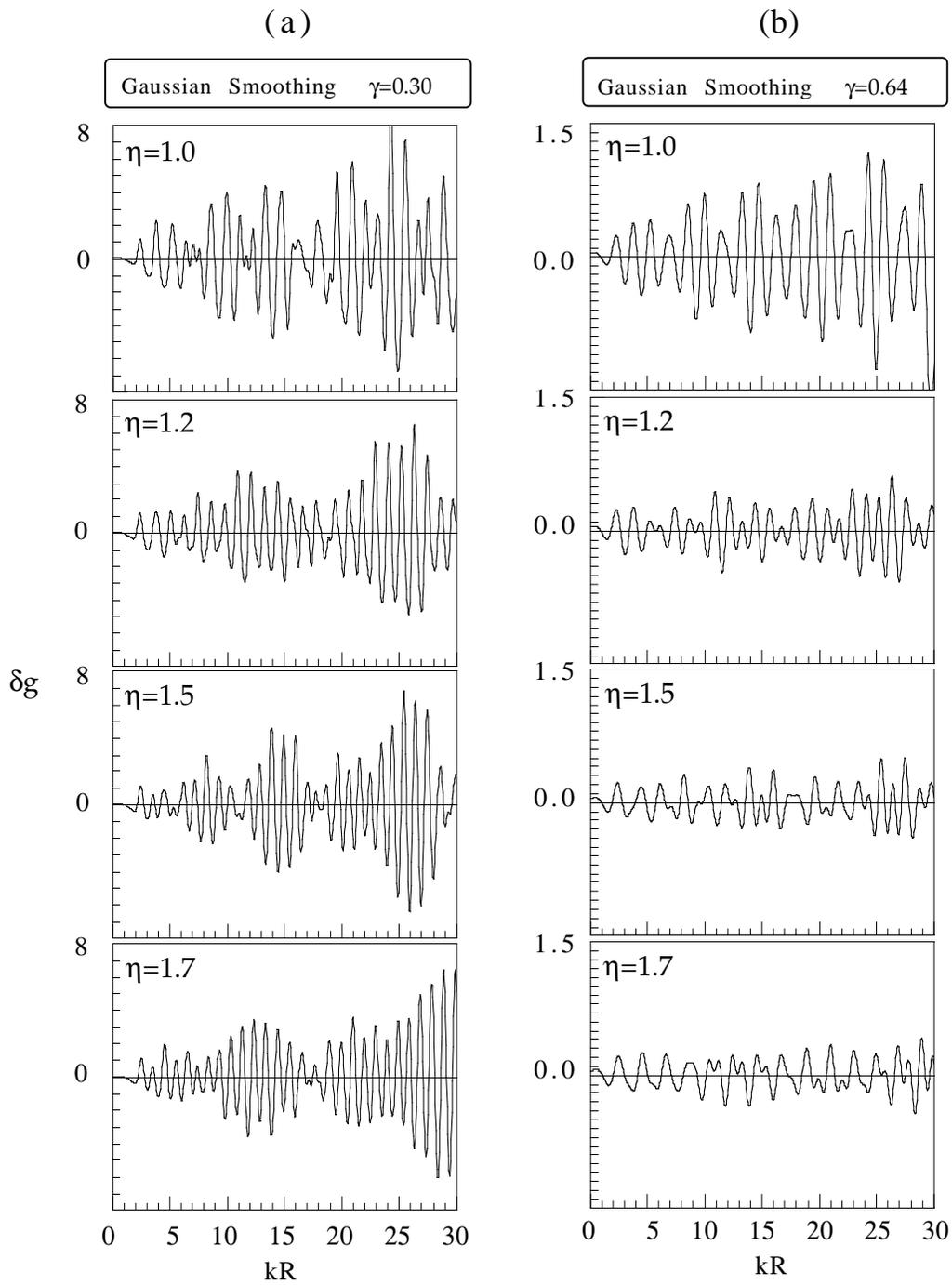}
\iffigs
\epsfxsize=.9\textwidth
\centerline{\epsffile{fig07.ps}}
\else
\vspace{184mm}
\fi
\caption{\label{fig7}
Coarse-grained level densities with the Gaussian smoothing parameter
$\gamma=0.3$ (a) and 0.64 (b).}
\end{figure}

With the aid of the Strutinsky averaging procedure,~\cite{st} clear
oscillatory patterns of the coarse-grained level density emerge as
shown in Fig.~\ref{fig7}, where (a) and (b) are obtained with the
Gaussian smoothing parameter $\gamma$ (defined by (\ref{gammas})) of
0.30 and 0.64, respectively.  As clearly seen from these figures, the
choice of a Gaussian smoothing parameter $\gamma$ is rather crucial for
properly identifying the coarse-grained level density, and hence the
contribution of classical periodic orbits.  At the circular limit
$\eta=1.0$, both Gaussian-smoothed level densities show similar
oscillations, whereas the shell gaps for $\gamma=0.64$ start to
collapse with increasing deformation.  In particular for deformations
$\eta$ larger than 1.5 the strong shell patterns cease to exist for the
case of $\gamma=0.64$, while for $\gamma=0.3$ the appreciable effects
still remain and show more oscillations as deformation increases.

In the semiclassical picture, for a given value of $\gamma$ the
contributions from only those periodic orbits of length up to $L_{\rm
max}\approx\pi R/\gamma$ can be considered.  In this context, it is
important to locate the actual shell-energy minima, irrespective of
the choice of a Gaussian smoothing parameter.

In terms of the particle number $N$ one can also obtain the
shell-correction energy $\delta E$ defined as the difference between
the sum of single-particle energies of $N$ lowest levels (taking the
spin-degeneracy factor 2 into account) and the Strutinsky averaged
energies, i.e.,
\begin{equation}\label{eq-shell}
\delta E=\sum_{i=1}^N\varepsilon_i-\tilde{E}, \qquad
\tilde{E}=2\int_{-\infty}^{\tilde{\varepsilon}_F}d\varepsilon'\,
\varepsilon'\,\tilde{g}(\varepsilon'),
\end{equation}
with the Fermi energy $\tilde{\varepsilon}_F$ satisfying
\begin{equation}\label{eq-num}
N=2\int_{-\infty}^{\tilde{\varepsilon}_F}d\varepsilon'\,
\tilde{g}(\varepsilon').
\end{equation}

\begin{figure}
\chkfig{fig08.ps}
\iffigs
\epsfxsize=.6\textwidth
\centerline{\epsffile{fig08.ps}}
\else
\vspace{108mm}
\fi
\caption{\label{fig8}
Shell structure energy $\delta E$ (in unit of $\varepsilon_0$)
plotted as a function of both
deformation $\eta$ and particle number $N$.}
\end{figure}

\begin{figure}
\chkfig{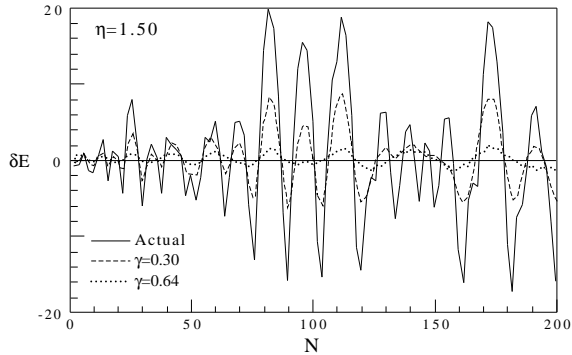}
\iffigs
\epsfxsize=.5\textwidth
\centerline{\epsffile{fig09.ps}}
\else
\vspace{50mm}
\fi
\caption{\label{fig9}
Smoothed shell-correction energies for $\eta=1.5$ with Gaussian
smoothing parameter $\gamma=0.3$ (dashed line) and 0.6 (dotted line).
Those without the smoothing are plotted by a solid line.}
\end{figure}

Figure~\ref{fig8} illustrates the oscillating pattern of the
shell-correction energy $\delta E$ as function of both deformation
$\eta$ and particle number $N$.  It is clear from the figure that the
distance between major shell gaps shrink with increasing deformation.
In the considered range of deformation it is found that the actual
magic numbers determined by the above procedure cannot be reproduced
with the choice of $\gamma=0.64$, whereas the value of $\gamma=0.3$ is
small enough not to demolish but still large enough to keep the actual
coarse-grained shell structure.  It is explicitly shown in
Fig.~\ref{fig9}, where the shell-correction energies are now calculated
by applying a Gaussian smoothing parameters of $\gamma=0.3$ and 0.64,
for the case of $\eta=1.5$ as an example.
In this case, the actual
magic numbers are found to be $\ldots,16,22,30,38,52,\ldots$, which
exactly coincide with those of $\gamma=0.3$, while those calculated
with $\gamma=0.64$ show larger oscillations missing
$\ldots,16,30,\ldots$.  The same is true for other deformations
considered in this paper. So, the coarse-grained shell structure
obtained with $\gamma=0.64$ is too rough and therefore we adapt
$\gamma=0.3$ to improve the precision of its description.

\subsection{Shell Structure and Fourier Spectra}
\label{qmlevdensf}

Equations of single-particle motion in the billiard are invariant with
respect to the scaling transformation
$(\boldr,\boldp,t)\to(\boldr,\alpha\boldp,\alpha^{-1}t)$.  The action
integral $S_\beta$ for a periodic orbit $\beta$ is proportional to its
length $L_\beta$,
\begin{equation}\label{scale}
S_\beta(E=p^2/2m) = \oint_\beta d\boldr\cdot\boldp
= pL_\beta = \hbar kL_\beta.
\end{equation}
and the semiclassical trace formula for the level density
is written as
\begin{equation}\label{eq-g}
g_{\rm scl}(\varepsilon) = \tilde{g}(\varepsilon)+\sum_{\beta}
A_{\beta}(kR) \cos\left(kL_{\beta}
-\frac{\pi}{2}\mu_{\beta}\right)
\end{equation}
where $\tilde{g}(\varepsilon)$ denotes the smooth part corresponding to
the contribution of zero-length orbits, $A_{\beta}=\left|{\cal
A}_{\beta}\right|$, $\mu_{\beta}$ the Maslov phase (the deformation and
energy dependent phase of Eqs.~(\ref{maslophasetot}) and
(\ref{maslovphasetotsd}) in our improved semiclassical approximation).
As previously discussed, the stationary phase approximation employed in
deriving the Gutzwiller trace formula breaks down at bifurcation points
for stable periodic orbits, and consequently results in the divergence
of the amplitudes $A_\beta(kR)$ in Eq.~(\ref{eq-g}), whereas in the
present ISP treatment those amplitudes are smooth functions of both
deformation and energy.

In order to examine the classical-quantum correspondence on shell
structure, one can perform the Fourier transform $F(L)$ of the quantum
level density $g(\varepsilon)$ with respect to the wave number $k$
\begin{equation}\label{eq-ft}
F(L)=\int dk\, e^{-ikL} g(\varepsilon) e^{-\frac{1}{2}
\left(\frac{ k}{k_c}\right)^2}
=\frac{1}{2\varepsilon_0R^2}\sum_i \frac{1}{k_i} e^{-ik_iL}e^{-\frac{1}{2}
\left(\frac{k_i}{k_c}\right)^2},
\end{equation}
which may be regarded as `length spectrum' exhibiting peaks at lengths
of individual periodic orbits. Here the Gaussian factor is imposed to
smoothly cutoff the spectrum in the high-energy region. In numerical
calculations, we use $k_c=k_{\rm max}/\sqrt2$, $k_{\rm max}$ being the
maximum wave number included. The above method of taking Fourier
transform of the quantum level density is known to be a powerful tool
to investigate the role of classical periodic orbits in the appearance
of shell fluctuations in quantum systems, and from such observations
one can also extract the semiclassical contributions of individual
periodic orbits.

\begin{figure}
\chkfig{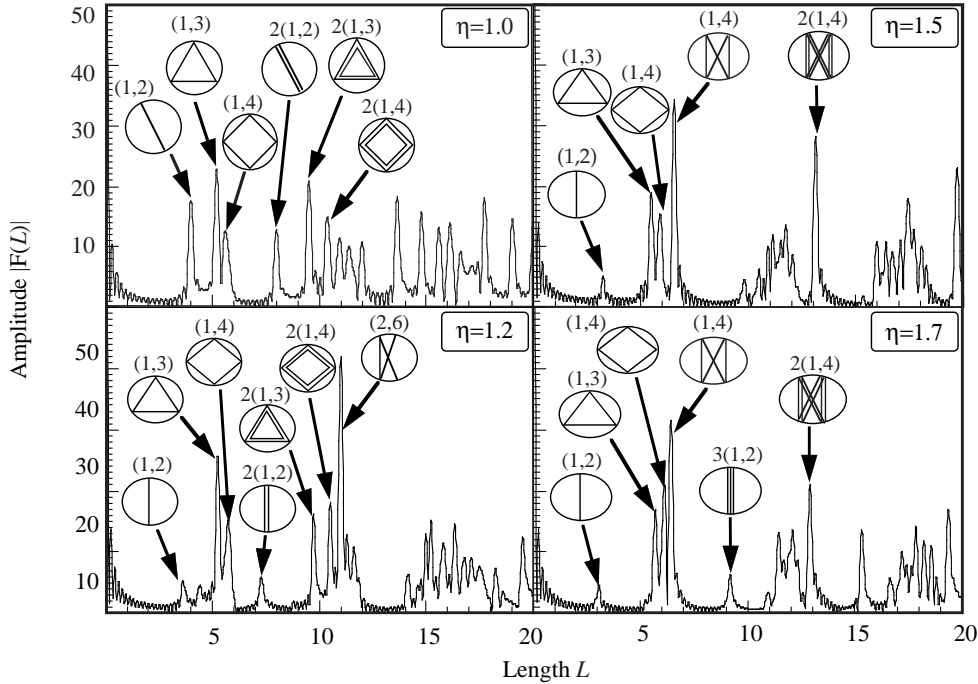}
\iffigs
\epsfxsize=.9\textwidth
\centerline{\epsffile{fig10.ps}}
\else
\vspace{97mm}
\fi
\caption{\label{fig10}
Fourier transforms of the single-particle level density for the
elliptic billiards with $\eta=1.0$ (a), 1.2 (b), 1.5 (c) and
1.7 (d). Some periodic orbits that correspond to peaks are illustrated.}
\end{figure}

Fourier spectra for deformations $\eta=1.0$, 1.2, 1.5 and 1.7 are
presented in Figs.~\ref{fig10}(a), (b), (c) and (d), respectively. At
axis ratio $\eta=1.0$, the diameter and elliptic orbits are found to be
equally important. The fact that only those shorter periodic orbits
mainly contribute to the gross-shell structure implies the significance
of three classical periodic orbits at the circular limit, namely the
diameter, triangular, and square shape orbits.  As deformation
increases, the Fourier amplitudes for triangular and rhombic orbits
still exhibit fairly strong effects, while those for diameter orbits
start to decline quickly and significant rearrangement can be observed.
Especially at deformations $\eta=1.5$ and 1.7, one can conclude, in
addition to triangular and rhombic shape orbits, the gross-shell
fluctuations are also governed by the (1,4) hyperbolic orbits
bifurcated from the 2(1,2) short diameter orbit at the critical
deformation $\eta=\sqrt{2}$.

\begin{sidewaysfigure}[p]
\chkfig{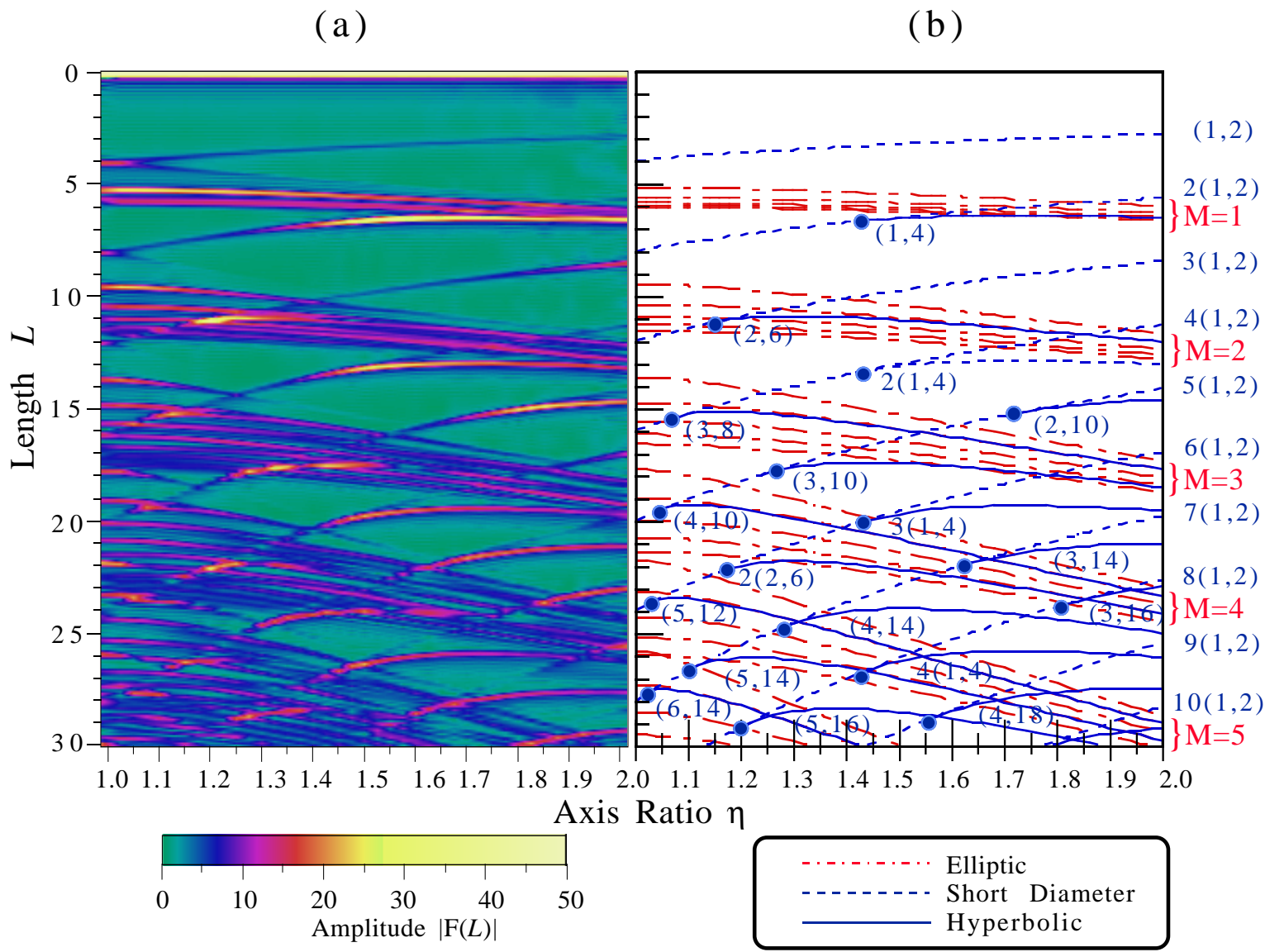}
\iffigs
\epsfxsize=.9\textheight
\centerline{\epsffile{fig11.ps}}
\else
\vspace{112mm}
\fi
\caption{\label{fig11}
(a) Modulus of the Fourier amplitudes plotted as functions of both
orbit-length $L$ and deformation $\eta$.
(b) Lengths $L$ of classical periodic orbits calculated as function
of deformation $\eta$.
Solid, dashed and dash-dotted lines are used for hyperbolic,
short-diameter, and elliptic orbits, respectively.}
\end{sidewaysfigure}

Figure \ref{fig11}(a) demonstrates the deformation dependence of
Fourier amplitudes calculated from the quantum single-particle
spectra.  Here the enhancement of peaks indicates a larger contribution
from the corresponding classical periodic orbits $\beta$ of length
$L_\beta$ to the shell structure.  At the circular limit the system
possesses the highest symmetry, and the breaking of this symmetry due
to a small deviation of its shape results in the orbital bifurcation.
With increasing deformation, the short diameter orbits with $M$
repetitions $M$(1,2) also bifurcate and create hyperbolic orbits at the
critical deformations $\eta_{\rm bif}$ given by Eq.~(\ref{etamin}).
The length of those classical periodic orbits as a function of
deformation can be calculated~\cite{nish90} as shown in
Fig.~\ref{fig11}(b).  It is clearly seen from both figures that the
bifurcations of stable periodic orbits give rise to an increase in the
Fourier amplitudes.  The remarkable enhancements seen in the figure
exactly coincide with the corresponding lengths of the newly created
hyperbolic orbits, and hence stress the importance of the orbital
bifurcations.

In this context, similar enhancements for the case of a spheroidal
cavity at superdeformed shape were also reported in
Ref.~\cite{ask}, where superdeformed shell structure is
associated with bifurcations of periodic orbits with two repetitions
on the equatorial plane.  In the present work, particular attention is
paid to investigate such correlations between bifurcations of stable
periodic orbits and quantum level-density oscillations.

\begin{figure}
\chkfig{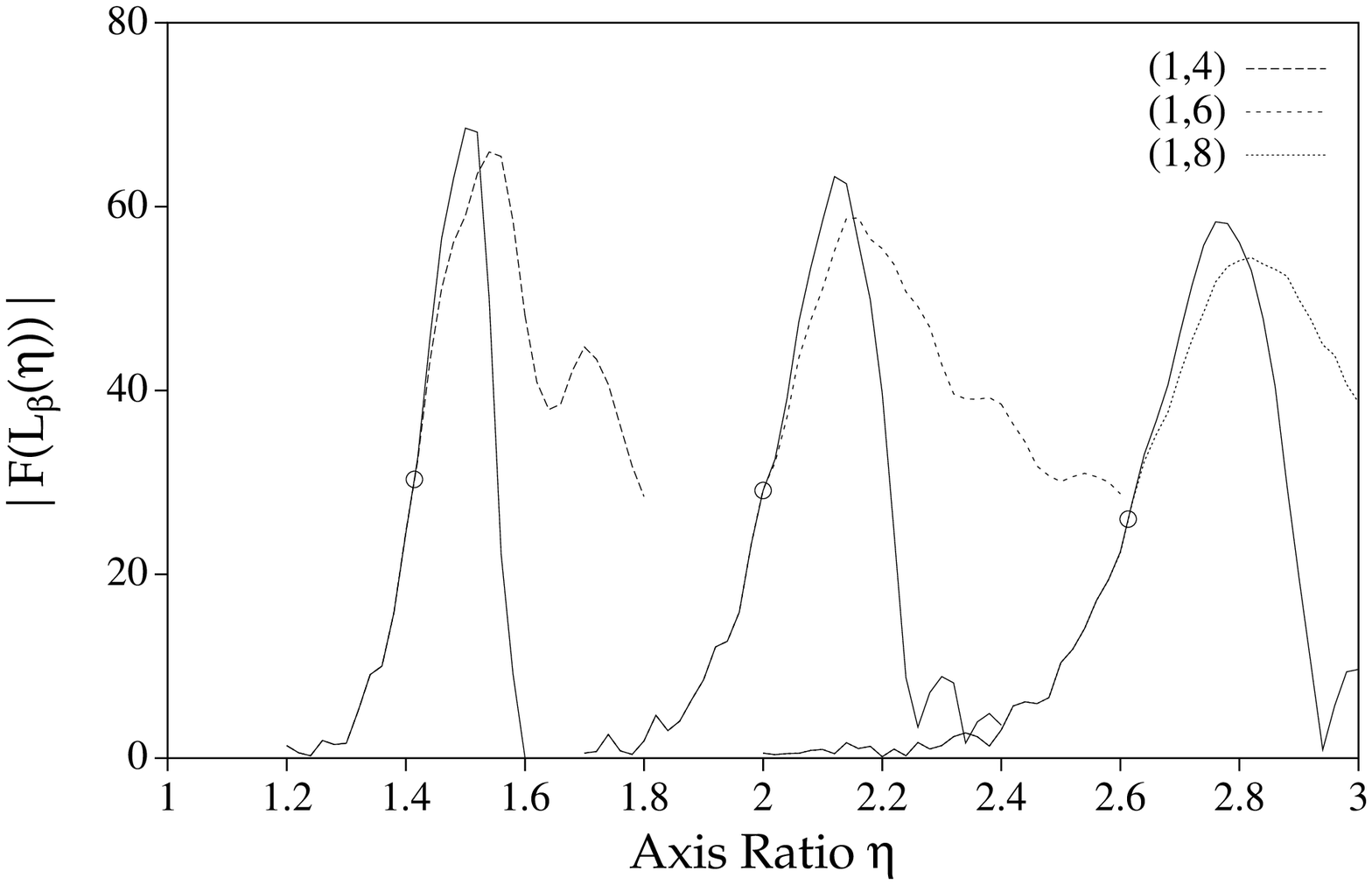}
\iffigs
\epsfxsize=.6\textwidth
\centerline{\epsffile{fig12.ps}}
\else
\vspace{61mm}
\fi
\caption{\label{fig12}
Deformation dependence of Fourier peak heights for hyperbolic and
short diametric orbits
2(1,2), (1,4), 3(1,2), (1,6), 4(1,2) and (1,8).
Solid lines are used for multiple traversals along the short diameter,
$M$(1,2) with $M=1,2,3$, while long-dashed, short-dashed and dotted lines
are used for hyperbolic (1,4),(1,6),(1,8) orbits, respectively.
Open circles indicate the bifurcation points.}
\end{figure}

In Fig.~\ref{fig12}, Fourier peak heights for some of the important
hyperbolic orbits, namely those bifurcated from the short diameter
orbits of 2, 3, and 4 repetitions, 2(1,2), 3(1,2) and 4(1,2), are
displayed as a function of deformation $\eta$. Interestingly, the
Fourier peaks for these newly created orbits show a rather universal
deformation dependence, that is, their heights reach the maxima shortly
after their bifurcation points and quickly decrease with increasing
deformation.  Such remarkable features were already seen in
Fig.~\ref{fig8}, where the shell valleys for $\eta$ approximately
larger than 1.5 can be understood to vary along the constant-action
lines $S(k,\eta)=\mbox{const.}$ of the (1,4) hyperbolic orbits, as
explained below.

Suppose some classical periodic orbits $\beta$ of length $L_\beta$ are
the dominant components in the semiclassical trace formula for the
oscillating level density, then the shell valley maxima/minima follows
the constant-action lines $S_\beta(k,\eta)=\mbox{const.}$ of those
dominating classical periodic trajectories.  Referring to
Eq.~(\ref{eq-g}), such lines are determined by
\begin{equation}\label{line}
kL_\beta-\frac{\pi}{2}\mu_\beta=(2n+1)\pi, \qquad n=0,1,2,\ldots
\end{equation}

\begin{sidewaysfigure}[p]
\chkfig{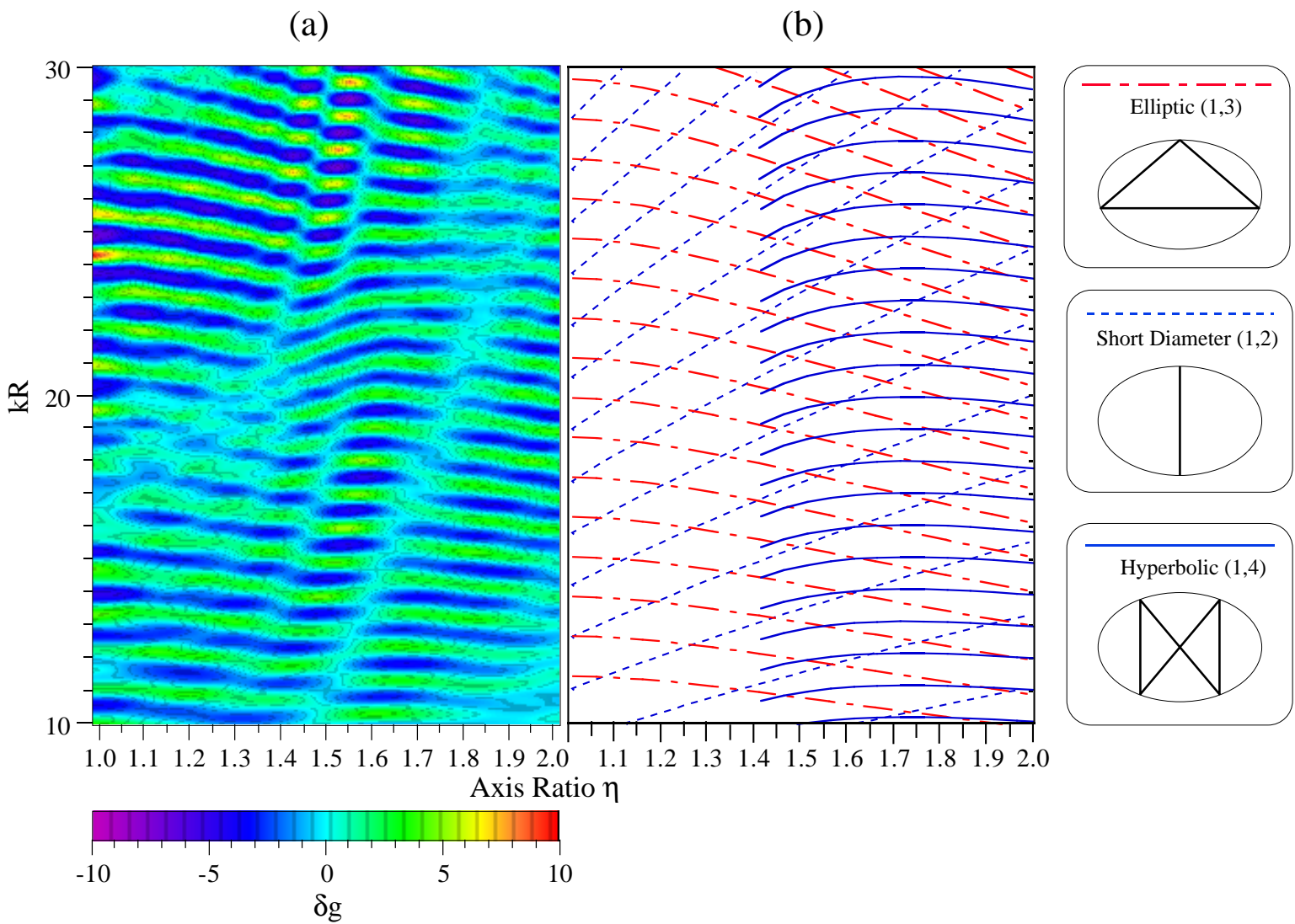}
\iffigs
\epsfxsize=.9\textheight
\centerline{\epsffile{fig13.ps}}
\else
\vspace{112mm}
\fi
\caption{\label{fig13}
(a) Smoothed level density plotted on the $k$-$\eta$ plane.
(b) Constant action lines on the $k$-$\eta$ plane for the elliptic
$(1,3)$ orbit (dash-dotted lines), the primitive short diameter $1(1,2)$
orbit (short-dashed lines) and the hyperbolic $(1,4)$ orbit (solid lines).}
\end{sidewaysfigure}

We shall now demonstrate the above dependence in Fig.~\ref{fig13}(a),
where the smoothed level densities are plotted on the $k$-$\eta$
plane.  As indicated in Fig.~\ref{fig13}(b), it is interesting to note
that the shell valley structures seen in Fig.~\ref{fig13}(a) can be
described by the constant-action lines of three major periodic orbits;
near the circular limit the shell valleys vary along those of elliptic
(mainly triangular and rhombic) orbits; in the right-half region of
Fig.~\ref{fig13}(a) the influence of newly created (1,4) hyperbolic
orbits is visible; the contribution of short diameter orbits are less
pronounced but certainly non-negligible throughout the considered range
of deformation.  The equality, Eq.~(\ref{line}), indicates the inverse
proportionality relation between the orbital length $L_{\beta}$ and
wave number $k$.  As the length of a trajectory $\beta$ increases, the
values of $k$ decrease and consequently the smoothed level densities
show more oscillations.  In particular, since the length of the (1,4)
hyperbolic orbits gradually increases within $\eta \approx
\sqrt{2}-1.7$ and then slowly decreases for $\eta>1.7$, the
corresponding constant-action lines also behave in the same manner.
Such a tendency was already observed in Fig.~\ref{fig8}, where the
contribution from the (1,4) hyperbolic orbits to the shell energy
$\delta E$ is apparent in the region $\eta>1.5$, indicating the
essential role of the orbital bifurcations in quantal shell formations.


\section{Comparison between Quantum and Semiclassical Calculations}

\begin{figure}
\chkfig{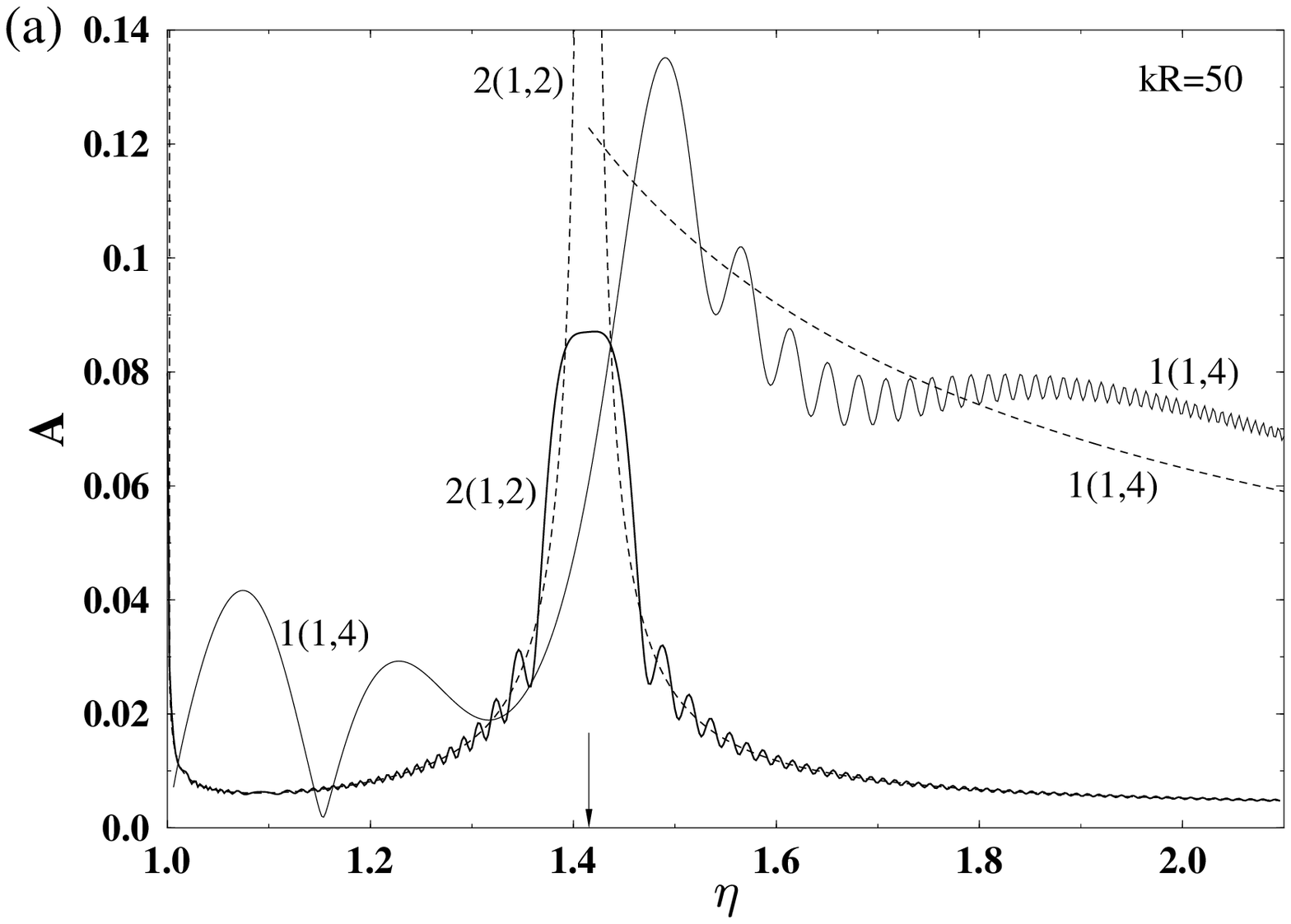}
\iffigs
\noindent
\begin{minipage}{.48\textwidth}
\epsfxsize=\textwidth\centerline{\epsffile{fig14a.ps}}
\end{minipage}\hfill
\begin{minipage}{.48\textwidth}
\epsfxsize=\textwidth\centerline{\epsffile{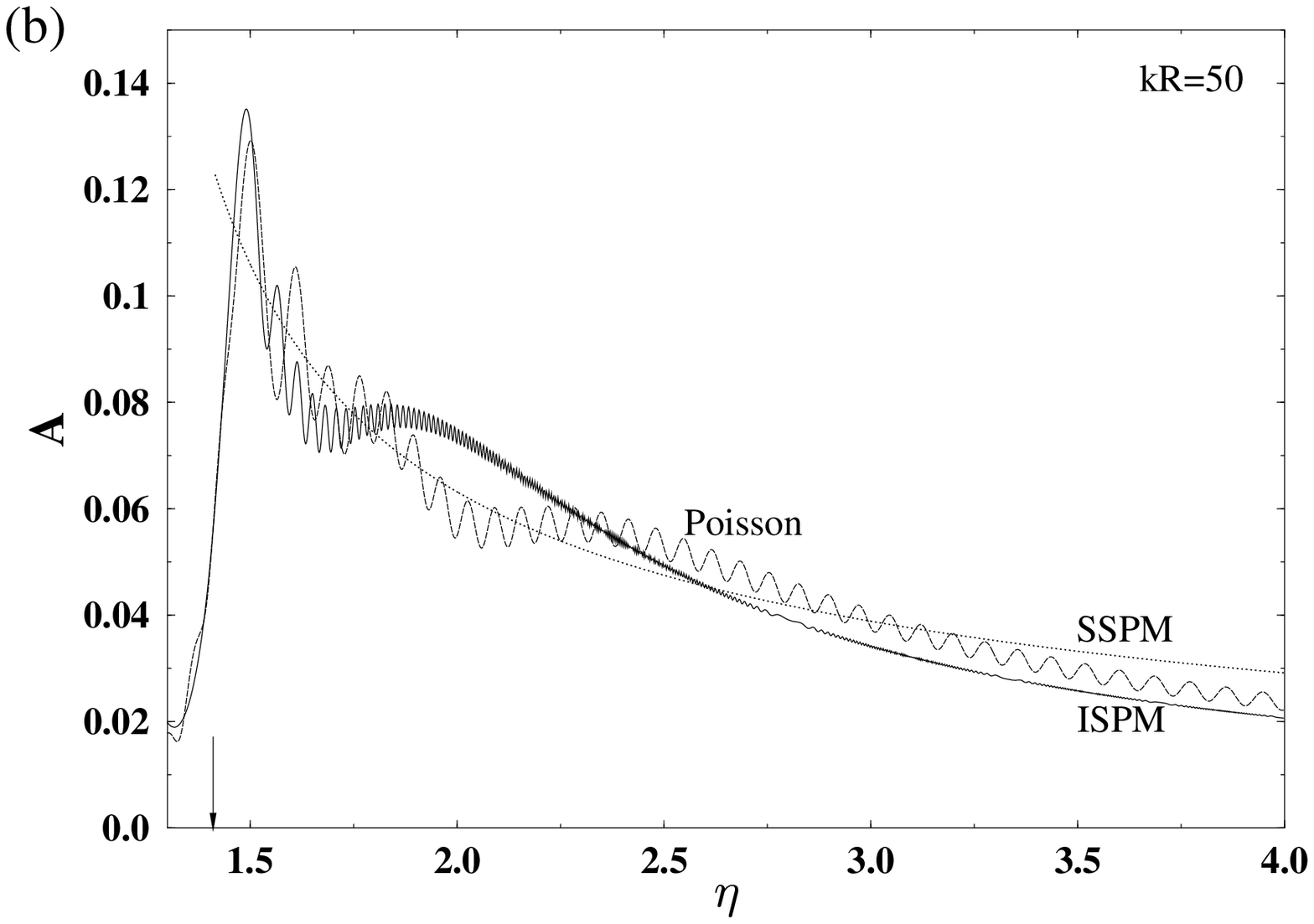}}
\end{minipage}
\else
\vspace{64mm}
\fi
\caption{\label{fig14}
(a) Amplitude modulus $A$
for bifurcating short diameter $2(1,2)$ and butterfly $1(1,4)$
orbits obtained by ISPM are shown by solid lines
as functions of deformation parameter $\eta$;
standard results of the extended Gutzwiller periodic orbit
theory (SSPM) are shown by short-dashed lines.
(b) ISPM amplitudes for the butterfly orbit (solid line)
are compared with exact calculation of the Poisson-sum
trace formula (\ref{poissonsum}) (dashed line marked by ``Poisson'')
and SSPM of Berry and Tabor~\cite{bt76} (dotted line).}
\end{figure}

\begin{figure}
\chkfig{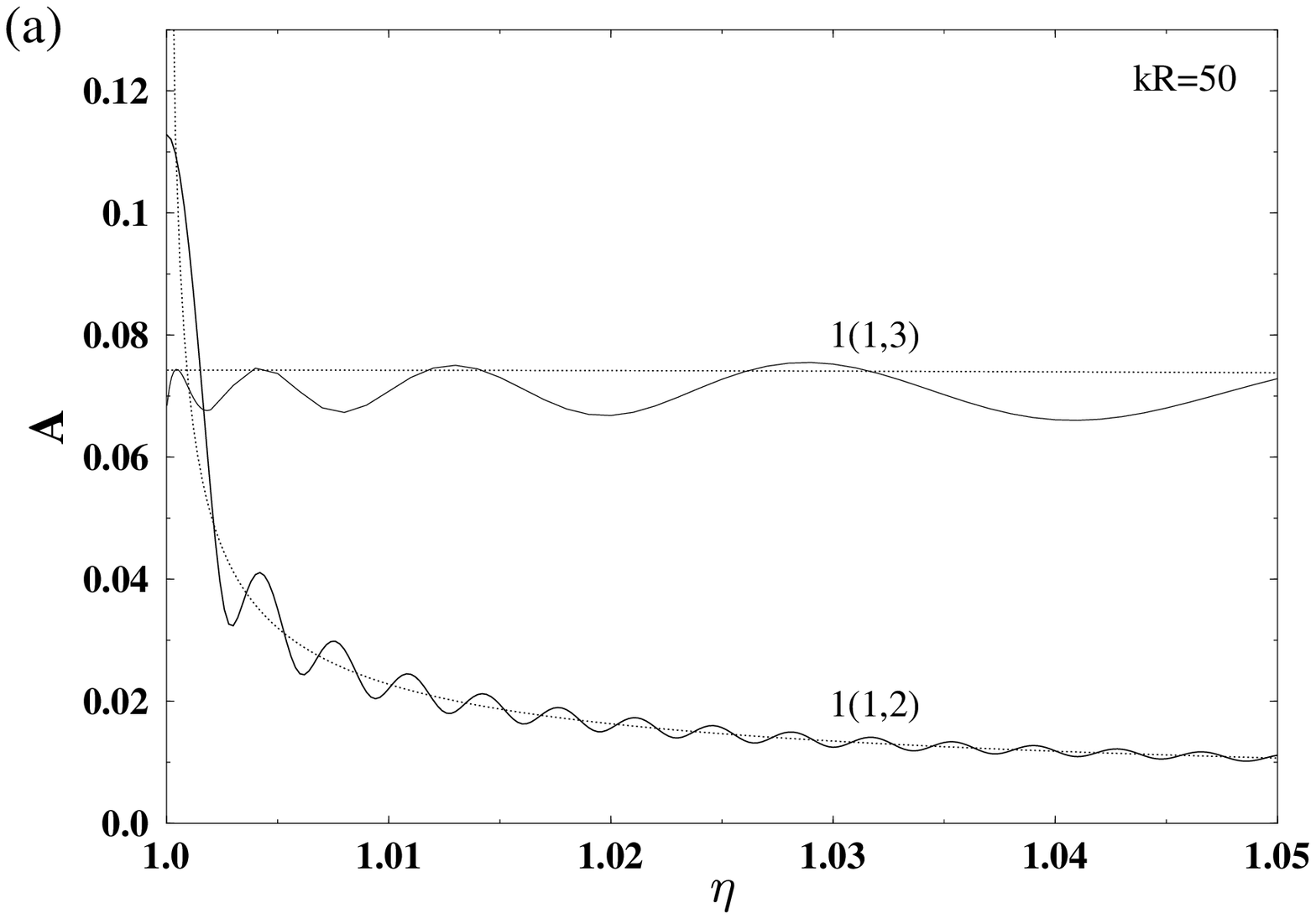}
\iffigs
\noindent
\begin{minipage}{.48\textwidth}
\epsfxsize=\textwidth\centerline{\epsffile{fig15a.ps}}
\end{minipage}\hfill
\begin{minipage}{.48\textwidth}
\epsfxsize=\textwidth\centerline{\epsffile{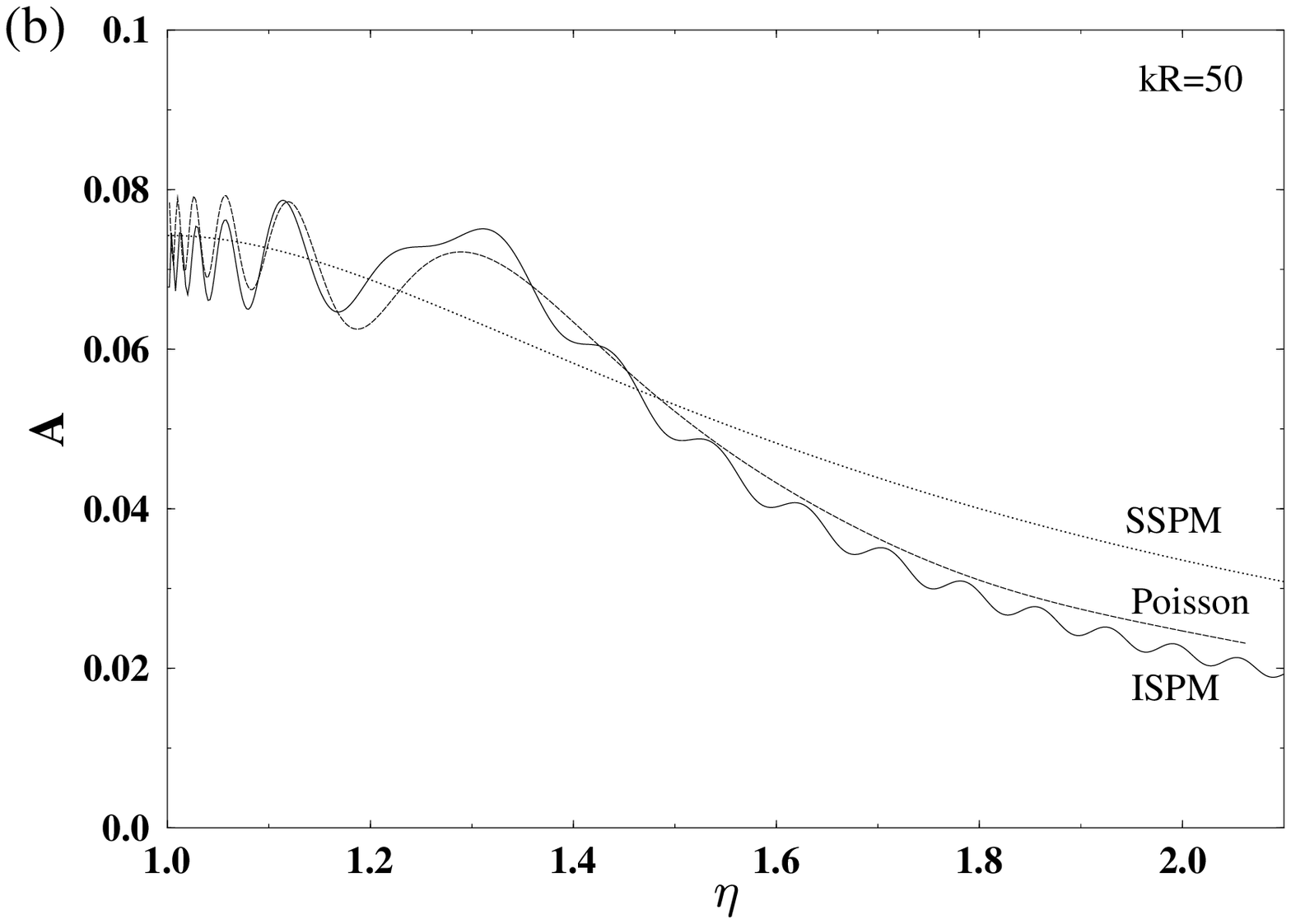}}
\end{minipage}
\else
\vspace{64mm}
\fi
\caption{\label{fig15}
(a) Same as in Fig.~\ref{fig14}(a) but
for primitive short diameter $1(1,2)$ and triangle $1(1,3)$ orbits
for smaller deformations.
(b) Comparison of the amplitudes for 1(1,3)
with exact calculations and SSPM.
Notations are the same as in Fig.~\ref{fig14}(b).}
\end{figure}

\begin{figure}
\chkfig{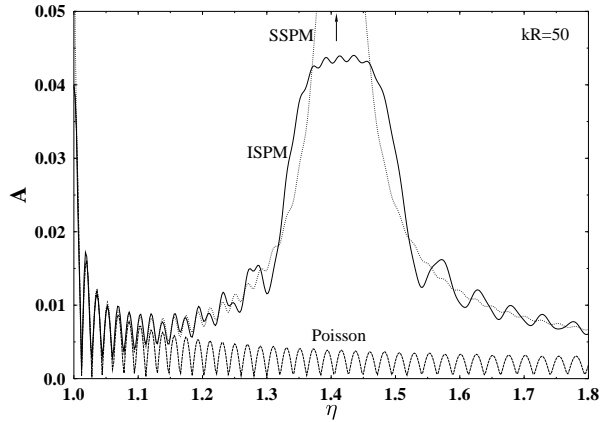}
\iffigs
\epsfxsize=.5\textwidth\centerline{\epsffile{fig16.ps}}
\else
\vspace{60mm}
\fi
\caption{\label{fig16}
ISPM amplitude modulus (solid line) for the sum of short
and long diameter $2(1,2)$ orbits is compared with the ($n_u=1,n_v=2,M=2$)
part of the Poisson-sum trace formula (\ref{poissonsum})
(long-dashed line) and the Gutzwiller SSPM (dotted line).}
\end{figure}

Figures~\ref{fig14},\ref{fig15},\ref{fig16} show the modulus of the
complex amplitude for a few short orbits.  The semiclassical amplitudes
for the hyperbolic ``butterfly'' $M(n_u,n_v)=(1,4)$ and elliptic
triangular (1,3) orbit families calculated by the ISPM are in good
agreement with the exact calculation of the Poisson-sum trace integral
(\ref{pstraceactang1}), see Figs.~\ref{fig14} and \ref{fig15},
respectively. All ISPM amplitudes are continuous function of the
deformation through the bifurcation point $\eta=\sqrt{2}$.  A
remarkable enhancement of the butterfly amplitude is seen at the
deformation $\eta=1.5\mbox{--}1.6$ slightly on the right of the
bifurcation point (see Fig.~\ref{fig14}).

The ISPM amplitude for the primitive short diameter 1(1,2) quickly
approaches the Gutzwiller SSPM result as one goes away from the
circular limit and, for larger deformations, its magnitude is
relatively small compared with those of other orbits mentioned above
(see Fig.~\ref{fig15}).

In Fig.~\ref{fig16} we compare the ISPM result with the modulus of
``diametric'' part of the Poisson-sum trace formula corresponding to
$n_u=1$, $n_v=2$ and $M=2$, which is regarded in Ref.~\cite{richens}
as to represent short and long diameters, as well as the standard
Gutzwiller results. The ISPM amplitude for the bifurcating short
diameter 2(1,2) has the two maxima; at the bifurcation deformation
$\sqrt{2}$, which is significantly larger than the butterfly and
triangular amplitudes, and at the circular shape, see also
Figs.~\ref{fig14} and \ref{fig15}. (Similar maxima at the circular
shape appear for any short diameter orbit. The maximum for the short
diameter 1(1,2) is the largest one, in particular, larger than for the
triangular orbit, see Fig.~\ref{fig15}(a).) As seen from
Fig.~\ref{fig16}, there is the same circular shape limit for the ISPM
approach and the ``diametric'' part of the Poisson-sum trace formula,
which is identical to the diameter family amplitude in the circular
disk.

Apparently, the behaviour of the ISPM amplitude for two repetitions of
the short diameter 2(1,2) is essentially different from that of the
``diametric'' part of the Poisson-sum trace integral which exhibits no
enhancement near the bifurcation point. Thus, the Poisson-sum trace
formula (\ref{poissonsum}) describes the families with maximum
degeneracy like hyperbolic and elliptic orbits, rather than the
isolated diameters. For the isolated orbits with smaller degeneracy
like diameters in elliptic billiard the Poisson-sum trace formula
cannot be applied because of the isolated stationary points for the
angle $\Theta_u$ variable.  This is the reason for the agreement of the
ISPM and SSPM asymptotics unlike for the ``diametric'' term of the
Poisson-sum trace integral in Eq.(\ref{poissonsum}). It means that the
diameters cannot be included in the usual EBK rational torus
quantization. However, the diameters could be included in a more
general quantization rule in terms of the averaged ISPM level densities
(\ref{tracetotal}) in a similar way as pointed out in
Refs.~\cite{strutmag,mfimb}.

We note a significant improvement of the ISPM results compared to the
SSPM for $\sigma$ close to the separatrix value $1$ and the creeping
value $\sigma_{\rm cr}$ (\ref{iucr}).  These cases might seem to be
important only in the limit $\eta \rightarrow \infty$ when $\sigma_{\rm
cr}$ tends to unity. However, even for $0\leq\eta\lsim 2$ we meet the
situations where the stationary points are close to the critical points
$\sigma=1$ and $\sigma=\sigma_{\rm cr}$, so that we have to integrate
within the finite limits.

\begin{figure}[p]
\chkfig{fig17a.ps}
\iffigs
\noindent
\begin{minipage}{.48\textwidth}
\epsfxsize=\textwidth\centerline{\epsffile{fig17a.ps}}
\end{minipage}\hfill
\begin{minipage}{.48\textwidth}
\epsfxsize=\textwidth\centerline{\epsffile{fig17d.ps}}
\end{minipage} \\
\begin{minipage}{.48\textwidth}
\epsfxsize=\textwidth\centerline{\epsffile{fig17b.ps}}
\end{minipage}\hfill
\begin{minipage}{.48\textwidth}
\epsfxsize=\textwidth\centerline{\epsffile{fig17e.ps}}
\end{minipage} \\
\begin{minipage}{.48\textwidth}
\epsfxsize=\textwidth\centerline{\epsffile{fig17c.ps}}
\end{minipage}
\else
\vspace{164mm}
\fi
\caption{\label{fig17}
Quantum and semiclassical (ISPM) oscillating level densities
$\delta g(kR)$ versus $kR$ for several deformations.
Averaging parameter $\gamma=0.3$, parameter of the
Strutinsky's shell correction method $\tilde{\gamma}=2.0$ and
correction polynomial degree $2{\cal M}=6$ are used.}
\end{figure}

We compare in Fig.~\ref{fig17} the semiclassical level densities
$\delta g_{\rm scl}(kR)$ calculated by the ISPM with the quantum
results for the averaging parameter $\gamma=0.3$.  The results obtained
by the ISPM are in good agreement with quantum results even near the
bifurcation point $\sqrt{2}$, where the SSPM gives the divergent result
due to the zeros of the stability factor $F_{sM}$ for the short
diameters 2(1,2).  For the deformations like $1.2$ and $1.7$ far from
the bifurcation, one obtains a fair agreement between the ISPM and the
SSPM.

\begin{figure}[p]
\chkfig{fig18a.ps}
\iffigs
\noindent
\begin{minipage}{.48\textwidth}
\epsfxsize=\textwidth\centerline{\epsffile{fig18a.ps}}
\end{minipage}\hfill
\begin{minipage}{.48\textwidth}
\epsfxsize=\textwidth\centerline{\epsffile{fig18b.ps}}
\end{minipage}
\else
\vspace{60mm}
\fi
\caption{\label{fig18}
Oscillating level density $\delta g(kR)$ versus $kR$ (left-hand side) and
shell energy $\delta E$ in units $\varepsilon_0$ versus $N^{1/2}$
(right-hand side) for a small deformation 1.01.
Solid and dotted lines indicate results of quantum and ISPM
calculations, respectively.
Parameters of the Strutinsky's shell correction
method are the same as in Fig.~\ref{fig17}.}
\bigskip

\chkfig{fig19.ps}
\iffigs
\epsfxsize=.75\textwidth\centerline{\epsffile{fig19.ps}}
\else
\vspace{108mm}
\fi
\caption{\label{fig19}
Quantum and ISPM shell energy $\delta E$ (in unit of $\varepsilon_0$)
are plotted by solid and dotted lines, respectively, as functions of
$N^{1/2}$.}
\end{figure}

Figure~\ref{fig18} shows a nice convergence of the ISPM results to
those of the circular disk trace formula for $\eta\to1$.  This
convergence is seen for any small deformation when the semiclassical
parameter $kR$ becomes sufficiently large.  With the inclusion of the
closed (periodic and non-periodic) hyperbolic orbit contribution, one
gets even better agreement with the quantum densities near the circular
disk shape.  For deformations far from the circular shape ($\eta \gsim
1.1$) and far from other bifurcation points, the contribution of the
hyperbolic ``co2'' orbits approaches Gutzwiller's SSPM result for the
isolated diameters, see Fig.~\ref{fig5}(b).

For the averaging parameter $\gamma=0.64$, we have good convergence of
POT sums for the ISPM and SSPM with a few short periodic orbits with
$M\leq1$, $n_u=1$ and $n_v\leq10$.  This is due to the damping factor
in Eq.~(\ref{dgsclgamma}) which ensures the convergence of the POT
sum.  For smaller $\gamma=0.3$ we need more orbits with $M\leq 2$,
$n_u\leq2$ and $n_v\leq10$.  Note that for $\gamma=0.3$ we have much
better agreement of the ISPM results with quantum mechanical
calculations than in the case of SSPM for the deformations near the
bifurcations including the transition to the circular shape, see
Fig.~\ref{fig7}.

\begin{sidewaysfigure}[p]
\chkfig{fig20.ps}
\iffigs
\epsfxsize=\textheight
\centerline{\epsffile{fig20.ps}}
\else
\vspace{112mm}
\fi
\caption{\label{fig20}
Shell energy maps $\delta E$ drawn as
function of $N^{1/2}$ and deformation $\eta$.
(Left-hand side) Quantum results like in
Figs.~\ref{fig16}, \ref{fig18} and \ref{fig19}:
(Middle- and right-hand sides)
Semiclassical ISPM results with and without taking into account
the bifurcating orbits, respectively (see text).}
\end{sidewaysfigure}

Figures 19 and 20 show nice agreements of the ISPM results for the
shell-correction energies with the corresponding quantum results. Note
that we substitute the exact Fermi energy $\varepsilon_F$ into the
semiclassical shell energy $\delta E$ (\ref{descl1}) by using the
second equation for the particle number there and quantum level density
like in Ref.~\cite{mfimbrk}.  This is important to get the correct
behaviour of the shell-correction energy as a function of particle
number $N$, as explained in Ref.~\cite{mfimbrk}. It is evident from
Fig.~\ref{fig20} that the nice agreement between the ISPM and quantum
results in the strongly deformed region of $\eta\geq\sqrt2$  cannot be
attained without including the contributions from bifurcating 2(1,2)
and (1,4) orbits.

In all our calculations we used the semiclassical approximation
improved at the bifurcation points which becomes better with increasing
$kR$ for all deformations including the bifurcation points.

\section{Conclusion}

The most essential new result of this paper in comparison to the
Berry-Tabor theory are two additional terms (second and third ones in
Eq.~(\ref{tracetotal})) in the improved trace formula for the elliptic
billiard. These two terms represent the contributions from the short
and long diameters which are continuous functions through all
bifurcation points. For deformations far from the bifurcation points,
we obtain asymptotically the standard Gutzwiller result for the
isolated diameters, and the correct trace formula for the diameters in
spherical limit of the circular billiard. Our results for the
hyperbolic and elliptic orbits improved near the bifurcation points are
simpler than those suggested within the uniform
approximation~\cite{bt76,bremen}.

With the use of our improved trace formula, we have demonstrated the
importance of bifurcations of the repeated short diameter orbit for the
emergence of shell structure at large deformations.

\section*{Acknowledgements}

We thank J. Blaschke, F. A. Ivanyuk, H. Koizumi, P. Meier,
V. V. Pashkevich, A. I. Sanzhur, and M. Sieber for many helpful
discussions.  Financial support by JSPS (grant No. RC39726005) and
INTAS (grant No. 93-0151) is gratefully acknowledged.


\appendix
\section*{Appendices}
\section{Curvatures}

The actions $I_u$ and $I_v$ given by Eq.~(\ref{actionuv}) are
expressed explicitly in terms of the elliptic
integrals~\cite{abramov,Byrd:Friedman}.  For elliptic orbits one has
\begin{eqnarray}
I_u &=& \frac{2}{\pi}\zeta\sqrt{2m\varepsilon\sigma}\,
\rE\left(\frac{\pi}{2},\frac{1}{\sqrt{\sigma}}\right),
\nonumber \\
I_v &=& \frac{1}{\pi}\zeta\sqrt{2m\varepsilon\sigma}
\left[\rE\left(\theta_e,\frac{1}{\sqrt{\sigma}}\right)
     -\rE\left(\frac{\pi}{2}, \frac{1}{\sqrt{\sigma}}\right)
     +\frac{\eta^2-\sigma(\eta^2-1)}{\eta \sqrt{\eta^2-1}}
\right].
\label{ensurfe}
\end{eqnarray}
For hyperbolic orbits,
\begin{eqnarray}
I_u &=& \frac{2}{\pi}\zeta\sqrt{2m\varepsilon}
\left[\rE\left(\frac{\pi}{2},\frac{1}{\sqrt{\sigma}}\right)
-(1-\sigma)\rF\left(\frac{\pi}{2},\frac{1}{\sqrt{\sigma}}\right)
\right], \nonumber \\
I_v &=& \frac{1}{\pi} \zeta\sqrt{2m\varepsilon}
\left\{(1-\sigma)\left[
\rF\left(\frac{\pi}{2},\frac{1}{\sqrt{\sigma}}\right)
-\rF\left(\theta_h,\frac{1}{\sqrt{\sigma}}\right)\right]
\right. \nonumber\\
&& \qquad\left.
+\rE\left(\theta_h,\frac{1}{\sqrt{\sigma}}\right)
-\rE\left(\frac{\pi}{2},\frac{1}{\sqrt{\sigma}}\right)
+\frac{\eta^2-\sigma(\eta^2-1)}{\eta\sqrt{\eta^2-1}}
\right\}
\label{ensurfh}
\end{eqnarray}
Eqs.~(\ref{ensurfe}) and (\ref{ensurfh}) may be regarded as equations
for the energy surface $\varepsilon(I_u,I_v)$ written in terms of the
parameter $\sigma$ for its elliptic and hyperbolic parts,
respectively.

The curvature $K$ of the energy curve are obtained by differentiating
Eqs.~(\ref{ensurfe}) and (\ref{ensurfh}) with respect to the parameter
$\sigma$. In this way one gets Eq.~(\ref{curvature}) with the
following derivatives for elliptic orbits,
\begin{eqnarray}
\frac{\partial I_u}{\partial\sigma}
&=& \frac{1}{\pi} \frac{\zeta\sqrt{2m\varepsilon}}{\sqrt{\sigma}}
\rF\left(\frac{\pi}{2}, \frac{1}{\sqrt{\sigma}}\right),
\nonumber \\
\frac{\partial^2 I_u}{\partial\sigma^2}
&=& -\frac{1}{2\pi} \frac{\zeta\sqrt{2m\varepsilon}}{\sqrt{\sigma^3}}
\Pi\left(\frac{\pi}{2},\frac{1}{\sigma},\frac{1}{\sqrt{\sigma}}\right),
\nonumber \\
\frac{\partial I_v}{\partial\sigma}
&=& -\frac{1}{2\pi} \frac{\zeta\sqrt{2m\varepsilon}}{\sqrt{\sigma}}
\left[\rF\left(\frac{\pi}{2},\frac{1}{\sqrt{\sigma}}\right)
-\rF\left(\theta_e,\frac{1}{\sqrt{\sigma}}\right)\right],
\nonumber \\
\frac{\partial^2 I_v}{\partial\sigma^2}
&=& \frac{1}{4\pi} \frac{\zeta\sqrt{2m\varepsilon}}{\sqrt{\sigma^3}}
\left[\Pi\left(\frac{\pi}{2},\frac{1}{\sigma},\frac{1}{\sqrt{\sigma}}
\right)-\Pi\left(\theta_e,\frac{1}{\sigma},\frac{1}{\sqrt{\sigma}}\right)
+\frac{\eta\sqrt{\eta^2-1}}{\sqrt{1-(1-\sigma^{-1})\eta^2}}\right].
\nonumber \\
\label{derivate}
\end{eqnarray}
For hyperbolic orbits,
\begin{eqnarray}
\frac{\partial I_u}{\partial \sigma}
&=& \frac{1}{\pi}\zeta\sqrt{2m\varepsilon}
\rF\left(\frac{\pi}{2},\sqrt{\sigma}\right), \nonumber\\
\frac{\partial^2 I_u}{\partial \sigma^2}
&=& \frac{1}{2\pi} \frac{\zeta\sqrt{2m\varepsilon}}{\sigma}
\left[\Pi\left(\frac{\pi}{2},\sigma,\sqrt{\sigma}\right)
-\rF\left(\frac{\pi}{2},\sqrt{\sigma}\right)\right], \nonumber\\
\frac{\partial I_v}{\partial \sigma}
&=& \frac{1}{2\pi}\zeta\sqrt{2m\varepsilon}
\left[\rF\left(\theta_h,\sqrt{\sigma}\right)
-\rF\left(\frac{\pi}{2},\sqrt{\sigma}\right)\right], \nonumber\\
\frac{\partial^2 I_v}{\partial \sigma^2}
&=& \frac{1}{4\pi} \frac{\zeta\sqrt{2m\varepsilon}}{\sigma}
\left[\Pi\left(\theta_h,\sigma,\sqrt{\sigma}\right)
-\Pi\left(\frac{\pi}{2},\sigma,\sqrt{\sigma}\right)
+\rF\left(\frac{\pi}{2},\sqrt{\sigma}\right)
-\rF\left(\theta_h,\sqrt{\sigma}\right)\right].
\label{derivath}
\end{eqnarray}

With Eq.~(\ref{derivate}) we obtain the curvature $K_\beta$
(\ref{curvature}) for elliptic orbits as
\begin{eqnarray}
K_\beta=\frac{\pi}{4p\zeta}
\frac{\kappa}{\rF^2(\frac{\pi}{2},\kappa)}
\left[\frac{\rF(\theta,\kappa)}{\rF(\frac{\pi}{2},\kappa)}
\Pi\left(\frac{\pi}{2},\kappa^2,\kappa\right)-\Pi(\theta,\kappa^2,\kappa)
+\frac{\eta\sqrt{\eta^2-1}}{\sqrt{1-(1-\kappa^2)\eta^2}}\right].
\nonumber\\
\label{curvate}
\end{eqnarray}
For hyperbolic orbits,
\begin{equation}
K_\beta=\frac{\pi}{4p\zeta}
\frac{1}{\kappa^2\rF^2(\frac{\pi}{2},\kappa)}
\left[\Pi(\theta,\kappa^2,\kappa)
-\frac{\rF(\theta,\kappa)}{\rF(\frac{\pi}{2},\kappa)}
\Pi\left(\frac{\pi}{2},\kappa^2,\kappa\right)\right].
\label{curvath}
\end{equation}

\section{Separatrix}

Like for the case of turning
points~\cite{fed:jvmp,masl,fed:spm,masl:fed} one writes
\begin{eqnarray}
\frac{1}{\hbar}\left[S_\alpha(\bbox{I}',\bbox{I}'',t_\alpha)
-(\bbox{I}''-\bbox{I}')\cdot\bbox{\Theta}''\right]
&=& c_0^\parallel+c_1^\parallel x+c_2^\parallel x^2+ c_3^\parallel x^3
+ \ldots \nonumber\\
&\equiv& \tau_0^\parallel+\tau_1^\parallel z +\frac13 z^3.
\label{expan3}
\end{eqnarray}
Here,
\begin{equation}
x=(I_u'-{I_u'}^*)/\hbar,
\label{x}
\end{equation}
\begin{eqnarray}
c_0^\parallel &=& \frac{1}{\hbar}\left[
S_\alpha^*(\bbox{I}',\bbox{I}'',t_{\alpha})
-(\bbox{I}'-\bbox{I}'')^*\cdot{\bbox{\Theta}''}^*\right]
=\frac{1}{\hbar}S_\alpha^*(\bbox{\Theta}',\bbox{\Theta}'',
\varepsilon),
\label{c0par} \\
c_1^\parallel &=& \left(\frac{\partial S_\alpha}{\partial I_u'}
-\Theta_u''\right)^*
=\Theta_u'-\Theta_u'' \to 0, \qquad (\sigma\to 1)
\label{c1par} \\
c_2^\parallel &=& \frac{\hbar}{2}\left(
\frac{\partial^2 S_\alpha}{\partial{I_u'}^2}\right)^*
=2\pi M\hbar K^\parallel\to\infty, \qquad (\sigma\to 1)
\label{c2par} \\
c_3^\parallel &=& \frac{\hbar^2}{6}\left(
\frac{\partial^3 S_\alpha}{\partial{I_u'}^3}\right)^*
=\frac{2\pi \hbar^2 M}{3}\left(
\frac{\partial K^\parallel}{\partial I_u}\right)<0,
\qquad (\sigma\to 1)
\label{c3par}
\end{eqnarray}
where the star means $I_u'=I_u''=I_u^*$.  The asymptotic behaviour of
the constants $c_i^\parallel$ near the separatrix $\sigma\approx 1$
was found from
\begin{equation}
K^\parallel\to \frac{\pi\log[(1+\sin\theta)/(1-\sin\theta)]}{
p\zeta(\sigma-1)\log^3(\sigma-1)}, \qquad (\sigma\to 1)
\label{Ksepasymp}
\end{equation}
$\theta\to\theta_h(\eta)$ formally, see (\ref{kappatheta}),
\begin{equation}
\frac{\partial K^\parallel}{\partial I_u}
\to -\frac{2\pi^2\log[(1+\sin\theta)/(1-\sin\theta)]}{
(p\zeta(\sigma-1)\log^2(\sigma-1))^2}. \qquad (\sigma\to 1)
\label{dKsepasymp}
\end{equation}
The second equality in Eq.~(\ref{expan3}) was obtained by a linear
transformation with some constants $\alpha$ and $\beta$,
\begin{eqnarray}
&x=\alpha z+\beta, \qquad
\alpha=(3 c_3^\parallel)^{-1/3}, \qquad
\beta=-c_2^\parallel/(3 c_3^\parallel),&
\label{alphabeta} \\
&\tau_0^\parallel=(c_0-c_1c_2/(3c_3)+2c_2^3/(27c_3^2))^\parallel, \qquad
\tau_1^\parallel=\alpha[c_1-c_2^2/(3c_3)]^\parallel.&
\label{taupar}
\end{eqnarray}
Near the stationary point for $\sigma\to 1$, one has
$c_1^\parallel\to 0$ and $\tau_1^\parallel\to -w_\parallel$ with the
positive quantity
\begin{equation}
w_\parallel=\left({c_2^2 \over {(3 c_3)^{4/3}}}\right)^\parallel
\to \left|\frac{M\log[(1+\sin\theta)/(1-\sin\theta)] p\zeta(\sigma-1)}{
2\hbar\log(\sigma-1)}\right|^{2/3}.
\label{wpar}
\end{equation}

Using expansion (\ref{expan3}) in Eq.~(\ref{pstraceactang}) and taking
the integral over angle $\Theta_v''$ exactly, i.e.  writing $2\pi$
instead of this integral, one gets
\begin{eqnarray}
\delta g_{\rm scl}^{(lM)}
&=& -\frac{2}{\hbar}\Re\sum_\alpha
\int d\Theta_u''\frac{1}{\left|\omega_v^*\right|}
e^{i(\tau_0-\nu_\alpha)}
\sqrt{\frac{\sqrt{w_\parallel}}{c_2^\parallel}}
\nonumber \\
&& \times
\left[\Ai\left(-w_\parallel,{\cal Z}_{lM,1}^\parallel,
{\cal Z}_{lM,2}^\parallel\right)
+i\Gi\left(-w_\parallel,{\cal Z}_{lM,1}^\parallel,
{\cal Z}_{lM,2}^\parallel\right)\right] \nonumber\\
&\approx& -\frac{2}{\hbar}\Re\sum_\alpha
\int d\Theta_u''\frac{1}{\left|\omega_v^*\right|}
e^{i(\tau_0-\nu_\alpha)}
\sqrt{\frac{\sqrt{w_\parallel}}{c_2^\parallel}}
\left[\Ai\left(-w_\parallel\right)
+i\Gi\left(-w_\parallel\right)\right],
\nonumber \\
\label{pstracelang}
\end{eqnarray}
where
\begin{equation}
{\cal Z}_{lM,1}^\parallel
= \sqrt{w_\parallel}, \qquad
{\cal Z}_{lM,2}^\parallel
= \sqrt{\frac{c_2^\parallel}{\sqrt{w_\parallel}}}\,
\frac{I_u^{\rm(cr)}}{\hbar}+\sqrt{w_\parallel}.
\label{zmimalpar}
\end{equation}
$\Ai(-w,z_1,z_2)$ and $\Gi(-w,z_1,z_2)$ are incomplete
Airy and Gairy functions,~\cite{frahn}
\begin{equation}
\left\{\begin{array}{c}
        \Ai(-w,z_1,z_2) \\
        \Gi(-w,z_1,z_2)
       \end{array}\right\}
= \frac{1}{\pi}\int_{z_1}^{z_2} dz\,
\left\{\begin{array}{c}
        \cos \\
        \sin
       \end{array}\right\}
\left(-wz+z^3/3\right),
\label{airynoncomp}
\end{equation}
and $\Ai(-w)$ and $\Gi(-w)$ are the corresponding standard
complete functions~\cite{abramov}.  We used in the second equation of
Eq.~(\ref{pstracelang}) that for any finite deformation $\eta$ and
large $kR$ near the separatrix ($\sigma\to 1$) one gets (see
Eq.~(\ref{wpar}))
\begin{eqnarray}
&&
{\cal Z}_{lM,1}^\parallel \to 0, \quad
{\cal Z}_{lM,2}^\parallel \to
4\left[\frac{M\log[(1+\sin\theta)/(1-\sin\theta)]p\zeta}{
2(\sigma-1)^2\log^4(\sigma-1)}\right]^{1/3} \nonumber\\
&&
\times\left[\frac{\eta}{\sqrt{\eta^2-1}}
\rE\left(\frac{\pi}{2},\frac{\sqrt{\eta^2-1}}{\eta}\right)-1
\right] \to\infty.
\label{asymptlim}
\end{eqnarray}

Using an analogous expansion of the action $\tau_0$ in
Eq.~(\ref{pstracelang}) with respect to the angle $\Theta_u''$ to the
third order and making a linear transformation like
Eq.~(\ref{alphabeta}), one arrives at Eq.~(\ref{deltagld}).  We
introduced in (\ref{deltagld}) several new quantities like
\begin{equation}
w_\perp=\left(\frac{c_2^2}{(3c_3)^{4/3}}\right)^\perp>0,
\label{wperp}
\end{equation}
\begin{equation}
{\cal Z}_{lM,2}^\perp=\sqrt{w_\perp}, \qquad
{\cal Z}_{lM,2}^\perp=\frac{\pi}{2}\left(
\left|3c_3^\perp\right|\right)^{1/3}+\sqrt{w_\perp},
\label{zmimalperp}
\end{equation}
\begin{equation}
c_2^\perp = \frac{1}{2\hbar}(J_\alpha^\perp)^*
= \frac{1}{2\hbar}\left(\frac{\partial^2 S_\alpha}{\partial{\Theta_u'}^2}
+2\frac{\partial^2 S_\alpha}{\partial\Theta_u'\partial\Theta_u''}
+\frac{\partial^2 S_\alpha}{\partial{\Theta_u''}^2}\right)_{lM}^*
= -\frac{F_{lM}}{8\pi MK^\parallel},
\label{c2perp}
\end{equation}
where $F_{lM}$ is the stability factor for long diameters,
see Eq.~(\ref{gutzstabfactl}),
\begin{eqnarray}
c_3^\perp &=& \frac{1}{6\hbar}
\left[\frac{\partial^3 S_\alpha}{\partial{\Theta_u'}^3}
+3\frac{\partial^3 S_\alpha}{\partial{\Theta_u'}^2\partial\Theta_u''}
+3\frac{\partial^3 S_\alpha}{\partial\Theta_u'\partial{\Theta_u''}^2}
+\frac{\partial^3 S_\alpha}{\partial{\Theta_u''}^3}\right]^*
\nonumber\\
&=& \frac{1}{6\hbar}
\left[\frac{\partial J_\alpha^\perp}{\partial \Theta_u'}
+\frac{\partial J_\alpha^\perp}{\partial\Theta_u''}\right]^*<0.
\label{c3perp}
\end{eqnarray}
Note, according to Eq.~(\ref{c2perp}), the quantity $c_2^\perp$ goes
to $0$ near the separatrix ($\sigma\to 1$) like for the caustic
case. This is the reason why the Maslov-Fedoryuk
theory~\cite{fed:jvmp,masl,fed:spm,masl:fed} can be used for the
transformation of the integral over angle $\Theta_u''$ in
Eqs.~(\ref{pstracelang}) into Eq.~(\ref{deltagld}).

\section{Jacobians for closed orbits with two reflection points}

The Jacobian $J_{\rm co2}^\parallel$ defined by the derivative in
Eq.~(\ref{jacobianalpha}) for closed orbits $\alpha$ like ``co2'' with
two reflection points,
$J_{\rm co2}^\parallel=
\left(\delta {\bar y}''/\delta\theta_p'\right)_{co2}$,
can be calculated by means
of the caustic method~\cite{mfimbrk}.
The main idea of this method is to use a specific property of the
trajectories in billiard system like elliptic cavity. They are the straight
lines which tangent a curve called
the elliptic or hyperbolic caustics
between turning points.
Our trajectory stability problem for the variations $\delta {\bar y}''$ at a
given $\delta \theta_p'$, see Fig.~\ref{fig3}, is much simplified
by reducing it to the calculation of the caustics semi-axes $a_c,b_c$ and
$a_c+\delta a_c,b_c+\delta b_c$ for closed orbit ``co2'' and
its $\delta \theta_p'$ deflection, respectively. For the case of closed
non-periodic orbits ``co2'' the semi-axes $a_c$ and $b_c$ and their variations
are functions of the initial point $(x,y)$ in contrast to the stability
problem for the periodic orbits of Ref.~\cite{mfimbrk}.  The orbit-length
invariant curve (confocal-to-boundary ellipse or hyperbola
crossing the point $(x,y)$, see Fig.~\ref{fig4})
and its semi-axis variations
play a similar role for the calculation of the ``co2'' stability
factor $J_{\rm co2}^\parallel$ with that of the boundary parameter
for the periodic orbits in Ref.~\cite{mfimbrk}.
In this way this stability factor is obtained in the form
\begin{equation}
J_{\rm co2}^\parallel=\frac{q_0-q_1}{\sqrt{1+q_1}}{\cal D},\qquad\qquad
{\cal D}=\frac{x''-x}{\delta \theta_p'}
\label{jacobco2}
\end{equation}
where $x''$ is the $x$-coordinate of the final point $O'$ (see
Fig.~\ref{fig3}),
$q_0$ and $q_1$ are the tangents of the slope angle for the initial and
final directions of particle motion along the orbit ``co2'',
\begin{equation}
q_0=\pm\frac{x_{c1}}{y_{c1}}\left(\frac{b_c}{a_c}\right)^2, \qquad
q_1=\pm\frac{x_{c2}}{y_{c2}}\left(\frac{b_c}{a_c}\right)^2,
\label{k01}
\end{equation}
Here, the upper and lower signs stand for the hyperbolic and elliptic closed
orbits, ($x_{c1},y_{c1}$) and ($x_{c2},y_{c2}$) are the first and last
tangent-to-caustics points of the trajectory ``co2'',
\begin{eqnarray}
x_{c1}&=&\frac{B_c+\sqrt{B_c^2-A_cC_c}}{A_c}, \quad
y_{c1} = \left\{\begin{array}{c}
                 1 \\
                 (a_c-x)/|a_c-x|
                          \end{array}\right\}
b_c \sqrt{1\pm\left(\frac{x_{c1}}{a_c}\right)^2},
\label{xyc1} \\
x_{c2}&=&\frac{B_c-\sqrt{B_c^2-A_cC_c}}{A_c}, \quad
y_{c2} = \left\{\begin{array}{c}
                 -A_c/|A_c| \\
                 1
                                        \end{array}\right\}
b_c \sqrt{1\pm\left(\frac{x_{c2}}{a_c}\right)^2},
\label{xyc2}
\end{eqnarray}
respectively, and
\begin{equation}
A_c=b_c^2x^2 \mp a_c^2y^2, \qquad
B_c=\mp a_c^2b_c^2 x, \qquad
C_c=a_c^4(b_c^2-y^2).
\label{ABC}
\end{equation}
The semi-axes $a_c$ and $b_c$ as functions of the initial
point ($x,y$) for the hyperbolic or elliptic caustics
for the orbit ``co2'' (see Fig.~\ref{fig4}) are given by
\begin{equation}
a_c = a\sqrt{\frac{\mp(b_x-b_c){\cal Z}}{b_x+b_c}}, \qquad
b_c = b\sqrt{1-{\cal Z}},
\label{abcaustics}
\end{equation}
where $a_x$ and $b_x$ are the semi-axes for the confocal-to-boundary
hyperbola or the ellipse crossing any current initial and final point
$(x,y)$ of the orbit ``co2'' inside the elliptic billiard,
\begin{eqnarray}
b_x^2 &=& \frac{x^2+y^2+b^2-a^2
\mp\sqrt{(x^2+y^2+b^2-a^2)^2-4y^2(b^2-a^2)}}{2}, \nonumber\\
a_x^2 &=& \mp(b_x^2-b^2+a^2),
\label{abcausticsx}
\end{eqnarray}
and ${\cal Z}$ is the root of the cubic algebraic equation,
\begin{eqnarray}
(1-\eta^2)^2{\cal Z}^3
&+&\left[(1+\eta^2)^2\left(\frac{b_x}{b}\right)^2+1-\eta^4
\right]{\cal Z}^2 \nonumber\\
&+&\left[2\eta^2-1-2(1+\eta^2)\left(\frac{b_x}{b}\right)^2
\right]{\cal Z}
+\left(\frac{b_x}{b}\right)^2-1=0.
\label{rootz}
\end{eqnarray}
The factor ${\cal D}$ in Eq.~(\ref{jacobco2}) is given by
\begin{equation}
{\cal D} = \frac{2a_x\Phi_a G}{A_0},
\label{D}
\end{equation}
where
\begin{equation}
\Phi_a = \eta^2 f_c
\left[\frac{\mp 4a_c^2b^2+\eta^2(a^2 \pm a_c^2)^2-b_c^4/\eta^2}{
2a_c(b^2-b_c^2 \pm \eta^2 a_c^2)^2}\right],
\label{fia}
\end{equation}
\begin{equation}
f_c = 2\left[d_0x+\frac{q_0(d_0^2-b^2+a^2)}{1+q_0^2}\right], \qquad
d_0 = y-q_0 x,
\label{fcd0}
\end{equation}
\begin{eqnarray}
G &=& \frac{2B_0d_0q_0+A_0 (b_x^2 \mp a_x^2-d_0^2)-C_0(1+q_0^2)}{
2\sqrt{B_0^2-A_0C_0}}-d_0q_0
\nonumber\\
&& +\frac{(1+q_0^2)(B_0-\sqrt{B_0^2-A_0C_0})}{A_0},
\label{G}
\end{eqnarray}
\begin{eqnarray}
A_0 = b_x^2 \mp a_x^2q_0^2, \qquad
B_0 = \mp a_x^2 d_0 q_0, \qquad
C_0 = \mp a_x^2(d_0^2-b_x^2),
\label{BC0}
\end{eqnarray}
We used here the invariance of the Jacobian ${\cal J}(x,y)$ against
time reversal.

\end{document}